\documentclass[aps,amsmath,amssymb,twocolumn,superscriptaddress]{revtex4}
  \usepackage{graphicx}
  \usepackage{color}
  \usepackage{enumitem}
  \usepackage{hyperref}

 \newcommand{\be}{\begin{equation}}
  \newcommand{\ee}{\end{equation}}
  
  \newcommand{\bea}{\begin{eqnarray}}
  \newcommand{\eea}{\end{eqnarray}}

  \newcommand{\p}{\partial}
  
  \newcommand{\la}{\left\langle}
  \newcommand{\ra}{\right\rangle}
  \newcommand{\lb}{\left[}
  \newcommand{\rb}{\right]}
  \newcommand{\lp}{\left(}
  \newcommand{\rp}{\right)}

  \renewcommand{\vec}[1]{{\boldsymbol #1}}
  \newcommand{\addLL}[1]{\textcolor{blue}{#1}}
  
  \newcommand{\addQ}[1]{\textcolor{red}{#1}}

\begin{document}

\begin{abstract} 
Achieving Bloch oscillations of free carriers under a direct current, a long-sought-after collective many-body behavior, has been challenging %notoriously hard 
due to stringent constraints on the band properties. We argue that the flat bands in moir\'e graphene fulfill the basic requirements for observing Bloch oscillations, offering an appealing alternative to the stacked quantum wells used in previous work aiming to %achieve 
access this regime. Bloch-oscillating moir\'e superlattices emit a comb-like spectrum of incommensurate frequencies, a property of interest for converting direct currents into high-frequency currents and developing broad-band amplifiers in THz domain. The oscillations can be synchronized through coupling to an oscillator mode in a photonic or plasmonic resonator. Phase-coherent collective oscillations in the resonant regime provide a realization of current-pumped THz lasing.
%features phase-coherent collective Bloch oscillations. 
%THz radiation.
%A detailed analysis of the phase diagram of the system %that 
%shows a transition between the asynchronous regime and a synchronized phase-coherent collective oscillation regime.
% Traveling wave amplification can be used to demonstrate gain. 
\end{abstract}

\title{Synchronizing 
Bloch-oscillating free carriers in moir\'e flat bands} % architectures} %by coupling to a single  oscillator mode}
\author{Ali Fahimniya${}^1$, Zhiyu Dong${}^1$, Egor I. Kiselev${}^2$, Leonid Levitov}
\affiliation{${}^1$Physics Department, 
Massachusetts Institute of Technology, Cambridge, Massachusetts 02139, USA\\
${}^2$Institut fur Theorie der Kondensierten Materie, Karlsruher Institut fur Technologie, 76131 Karlsruhe, Germany}
    
\maketitle  
Bloch oscillations, arising when electrons are driven through a perfect crystal lattice by an electric field, are an iconic example of a coherent dynamics in %externally driven 
quantum many-body systems\cite{ashcroft_mermin,pippard}. 
The oscillations are at the same frequency for all carriers, for a one-dimensional lattice %the frequency is 
given by $\omega=eEa/\hbar$ with $E$ the field strength and $a$ the lattice period. 
%governed %determined solely 
%by lattice periodicity $a$ and field strength $E$, so that all carriers oscillate at the same frequency, for a one-dimensional lattice given by $\omega/2\pi=eEa/h$. 
Besides the obvious fundamental appeal, this behavior 
has long been eyed as a promising way % of interest as a way
%points to  an appealing possibility 
to convert direct currents into high-frequency currents\cite{Esaki1970}.  
Wide interest in this phenomenon stems from the expectation that it may 
%Bloch-oscillating electrons have the potential to 
%provide broad band gain at THz frequencies\cite{Ktitorov1972,Kroemer2000} and 
%become the basis of a technology that will 
help fill the infamous ``THz gap'', % in solid state THz fundamental oscillators, 
leading to radiation emitters and detectors operating in %the THz 
this frequency range\cite{Ktitorov1972,Kroemer2000,Savvidis2004}. 

% and, therefore, \addLL{can give rise to sharp resonances in the radiation power spectrum\cite{ashcroft_mermin,pippard}.} 
%\addLL{Fundamental interest?}

While Bloch oscillations have long been immortalized in textbooks, realizing them in solids has proven to be %notoriously difficult. 
a challenging task. % due to 
%was impeded by fairly stringent requirements on the system.
Achieving this regime requires overcoming several obstacles. One is the dephasing due to electron energy loss to phonons. % emission that dephases the oscillations. 
To suppress phonon emission exceptionally narrow electronic bands of width smaller than the optical phonon energy must be used. Another is the dephasing due to disorder scattering. Experimental efforts so far mainly focused on narrow minibands in synthetic MBE-grown semiconductor superlattices\cite{Feldmann1992,Waschke1993,Sibille1990,Savvidis2004}.
%, which host tunable narrow minibands. %of narrow width 
%tunable by the superlattice design parameters\cite{Feldmann1992,Waschke1993,Sibille1990,Savvidis2004}. 
These %superlattice 
systems cleared a number of key milestones on the road towards achieving Bloch oscillations. %Namely, 
They display the signatures indicative of Bloch oscillations such as negative differential conductivity $dI/dV<0$, recurrence and ringing in the optical pump-probe measurements, Wannier-Stark (WS) ladders and, last but not least, optical gain\cite{Feldmann1992,Waschke1993,Sibille1990,Savvidis2004}. However, upon the injection current %increasing and 
approaching the relevant parameter range the superlattice systems develop instabilities and show a complex noisy behavior due to the onset of switching and formation of electric domains. This behavior presents the main obstacle to achieving the collective globally-synchronized Bloch oscillations\cite{Hyart2008,Hyart2009a,Hyart2009b}. 

Meanwhile, recently Bloch oscillations were achieved in cold atom systems, using Bloch minibands in optical lattices\cite{Dahan1996,Anderson1998,Morsch2001,Cristiani2002,Ott2004}. 
%Because the cold-atom systems are charge-neutral, instead of the electric field the force of gravity had to be used to accelerate particles. 
This proof-of-principle demonstration has greatly improved our understanding of the underlying physics\cite{Gluck02,Kolovsky2013} and strengthened interest in demonstrating electronic Bloch oscillations. 

Given the difficulties encountered in semiconducting superlattices it is natural to seek %look for 
other systems that meet the requirements for achieving Bloch oscillations.
%, but are not prone to the electric domain formation instabilities. 
%One appealing possibility is to employ 
One enticing opportunity is offered by the recently introduced moir\'e superlattices in twisted bilayer graphene, a material that hosts electron bands that are tunable by the twist angle\cite{Bistritzer2011,Cao2018a,Cao2018b,Cao2016,Kim2016,Berdyugin2020}.  For twist angles $\theta\lesssim 2^\circ$ the moire electron bands are considerably narrower than the optical phonon energy ($\sim 200$ meV), becoming as narrow as $J\lesssim 10$-$20$ meV near ``magic" values of the twist angle $\theta\sim 1^\circ$. Such bandwidths are sufficient to eliminate the optical phonon emission, the main obstacle to observing coherent Bloch oscillations in wide bands.  

The moir\'e graphene also clears other key requirements for observing Bloch oscillations. One is weak disorder scattering. Since the narrow bands are formed in a solid with a pristine near-perfect atomic order, they are less susceptible to disorder than the bands in synthetic MBE-grown semiconductor superlattices. This is manifested in a high carrier mobility and ballistic carrier transport observed over micron lengthscales at $T=0$\cite{Kim2016,Berdyugin2020}. Estimating the scattering time as $\tau=l/v_F$ with the mean free path $l\sim 1\,{\rm \mu m}$ and velocity $v_F$ of about $1/30$ of the graphene monolayer value $10^6\,{\rm m/s}$ gives $\tau\sim 3\cdot 10^{-11}$s, a value comparable to that of graphene monolayer. The scattering rate can therefore be as low as %two orders of magnitude smaller than the bandwidth, 
$\gamma_{\rm dis} \sim 10^{-2}J$. 
%Further, %more, %it is all but natural to expect that 
The two-dimensional character of moir\'e graphene will also help to suppress the instability 
towards the formation of electric field domains that hindered experiments in the stacks of quantum wells\cite{Savvidis2004}. 
%due to inhomogeneous charge accumulation that drives the formation of electric field domains occurring in the stacks of quantum wells. 
%Indeed, in 
In the moir\'e setup the electric current %is 
can be driven
in the graphene plane in a manner that maintains the translation invariance of the system 
and does not cause local charging. %In addition, 
Indeed, gating is known to maintain a spatially uniform carrier density
even under moderate to high currents. 

Other appealing properties of moir\'e graphene %an appealing platform to demonstrate Bloch oscillations 
are the lack of Zener transitions, % due to %relatively large size of the 
suppressed by sizable minigaps, and %separating 
% above and below the flat bands, and  % from %other 
%the higher and lower minibands, and 
the weakness of the electron %scattering by 
coupling to the long-wavelength 
acoustic phonons\cite{Bistritzer2009,Tse2009,Song2012}. %Added to that,  
Further, the relatively large periodicity of moir\'e superlattices ($a\sim 10$nm) reduces the required $E$ field values: 
\be
\gamma={\rm max}[\gamma_{\rm ph},\gamma_{\rm dis}] < \omega_{\rm B} < J/\hbar,
\quad 
\omega_{\rm B} = eEa/\hbar
.
\ee 
Using moderate $E$ fields will help to avoid the %Wannier-Stark ladder 
WS localization effects and charge instabilities. 

A key assumption is that phonon emission can remain relatively weak despite rate enhancement due to the high density of states in moir´e bands and an out-of-equilibrium carrier state created under an applied current. These expectations are supported by a detailed analysis of phonon emission\cite{Supplementary Information}, %in %carrier dephasing 
%moir\'e bands\cite{Supplementary Information}, % of phonon emission in twisted bilayer graphene, 
predicting %a drop in 
emission rates which drop upon an increase in the flat-band width and a growing $E$ field. Detuning away from the magic twist angle reduces the density of states that govern phonon emission. Likewise, an $E$ field tunes the % Wannier-Stark
WS states out of resonance, abruptly quenching phonon emission.
% and enhancing coherence of Bloch-oscillating carriers. 

\begin{figure}[t]
\includegraphics[width=0.99\columnwidth]{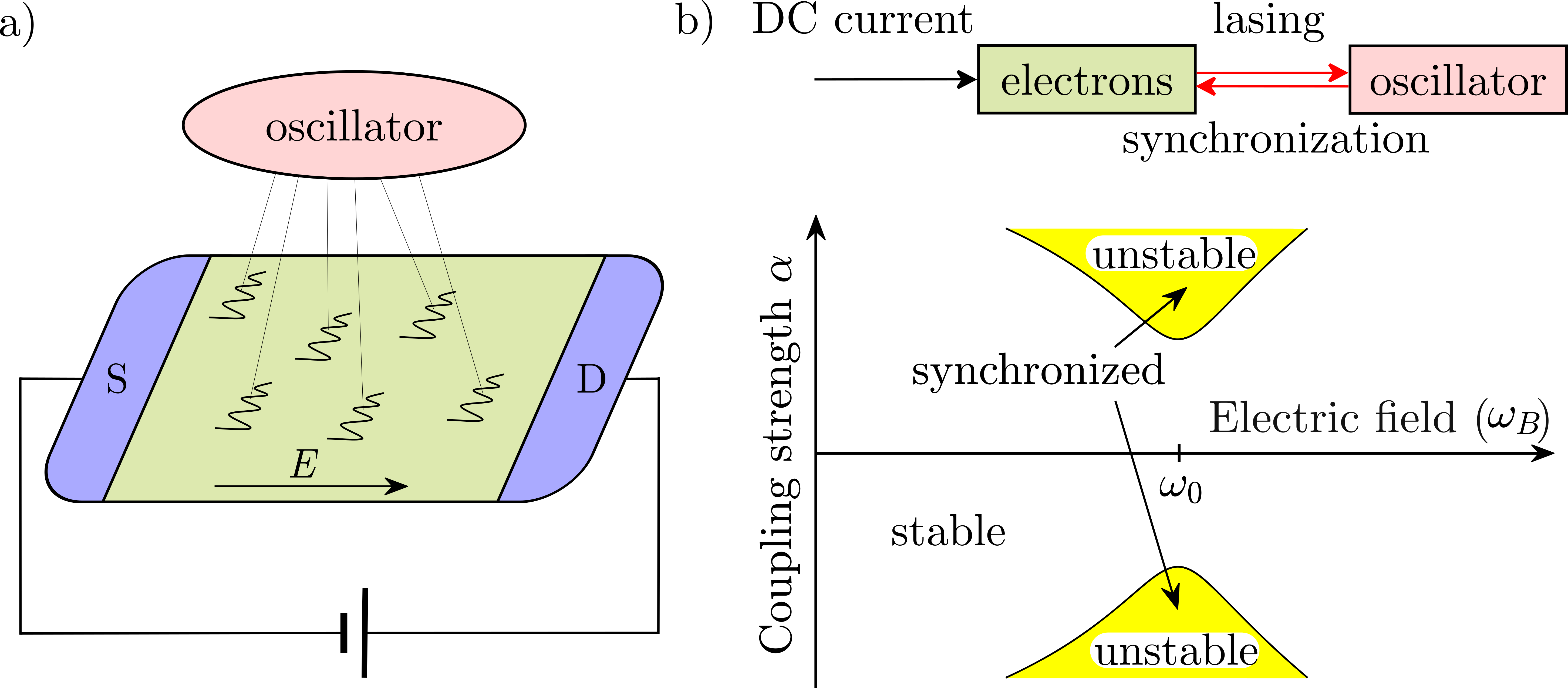}
\caption{
a) Bloch-oscillating electrons synchronized by coupling to an oscillator mode. 
A DC electric field $\vec E$ %due to voltage source 
drives 
free-carrier oscillations with frequency $\omega_{\rm B}$ (wavy lines). 
The oscillations are at the same frequency for all carriers but are asynchronous (not in phase). 
Synchronized oscillations are achieved through coupling to an oscillator mode, depicted by the thin lines.
%The couplings between the electrons and the oscillator mode, depicted by the thin lines, can synchronize the oscillations. 
b) A phase diagram showing the stable and unstable regimes, in which Bloch oscillations are asynchronous and synchronized, respectively. 
The carrier scattering 
rate $\gamma$ is taken to be equal the oscillator damping %rate 
$\gamma_0$ (see Eq.\eqref{eq:characteristic_with_gamma0}); phase diagrams for unequal $\gamma$ and $\gamma_0$ are discussed in \cite{Supplementary Information}. %Online Supplement. %pictured in Fig.\ref{fig2_oscillator}. 
The Bloch frequency $\omega_{\rm B}$ on the $x$ axis is proportional to the electric field; $\omega_0$ is the 
oscillator frequency,  
the coupling strength $\alpha$ between electrons and the oscillator is defined in Eq.\eqref{eq:EoM_singleMode}. 
Instability is easiest to achieve when $\omega_{\rm B}$ is tuned close to $\omega_0$. The flowchart on top shows the relationship between different degrees of freedom: %the electrons in the system and the oscillator mode in the unstable phase. T
the DC current 
drives free-carrier oscillations; these, synchronized by the oscillator, 
%, the oscillations are synchronized by the oscillator and, in turn, 
pump energy into it (the lasing effect). 
%\addQ{[explain lasing, mention figure early]}
%\addQ{[Does caption describe figure step by step? Anything to add?]}
}
\label{fig1_oscillator}
%\vspace{-5mm}
\end{figure}
% https://docs.google.com/drawings/d/1x_BV_mlNFzWE0tP6zw4nGRe0idDlnBe5OcOYBcw3PJk/edit?ts=5eef84b9

Importantly, although all free carriers Bloch-oscillate with identical frequencies, these oscillations are \emph{asynchronous,} as %Namely, under a DC current, 
the oscillation phases are totally random and uncorrelated for different carriers. Therefore, in order to achieve collective continuous-wave Bloch oscillations driven by a direct current, %the movement of different carrier 
movements of different carriers must be synchronized. We outline a way to achieve this %discuss how this can be achieved 
through coupling of the current-carrying channel to an oscillator mode in a THz resonator. The resonator frequency depends on system parameters, %at the same time 
whereas the Bloch frequency is tunable by varying the applied electric field. %As we will see, 
As illustrated in Fig.\ref{fig1_oscillator}, this system develops an instability towards collective oscillations at a Bloch frequency when the latter is close to the oscillator frequency. In practice, the oscillator can be realized as a THz photonic or plasmonic resonator in a 2D or a 3D architecture\cite{Savvidis2004,Ju2011,Yan2012a,Yan2012b,Tu2020,Ateshian2020}. 
An alternative route to achieve synchronization is through coupling to an intrinsic collective mode, excitonic or plasmonic. 
Phase-coherent oscillations achieved in this regime represent a realization of electrically pumped THz lasing.

\begin{figure}[t]
\includegraphics[width=0.99\columnwidth]{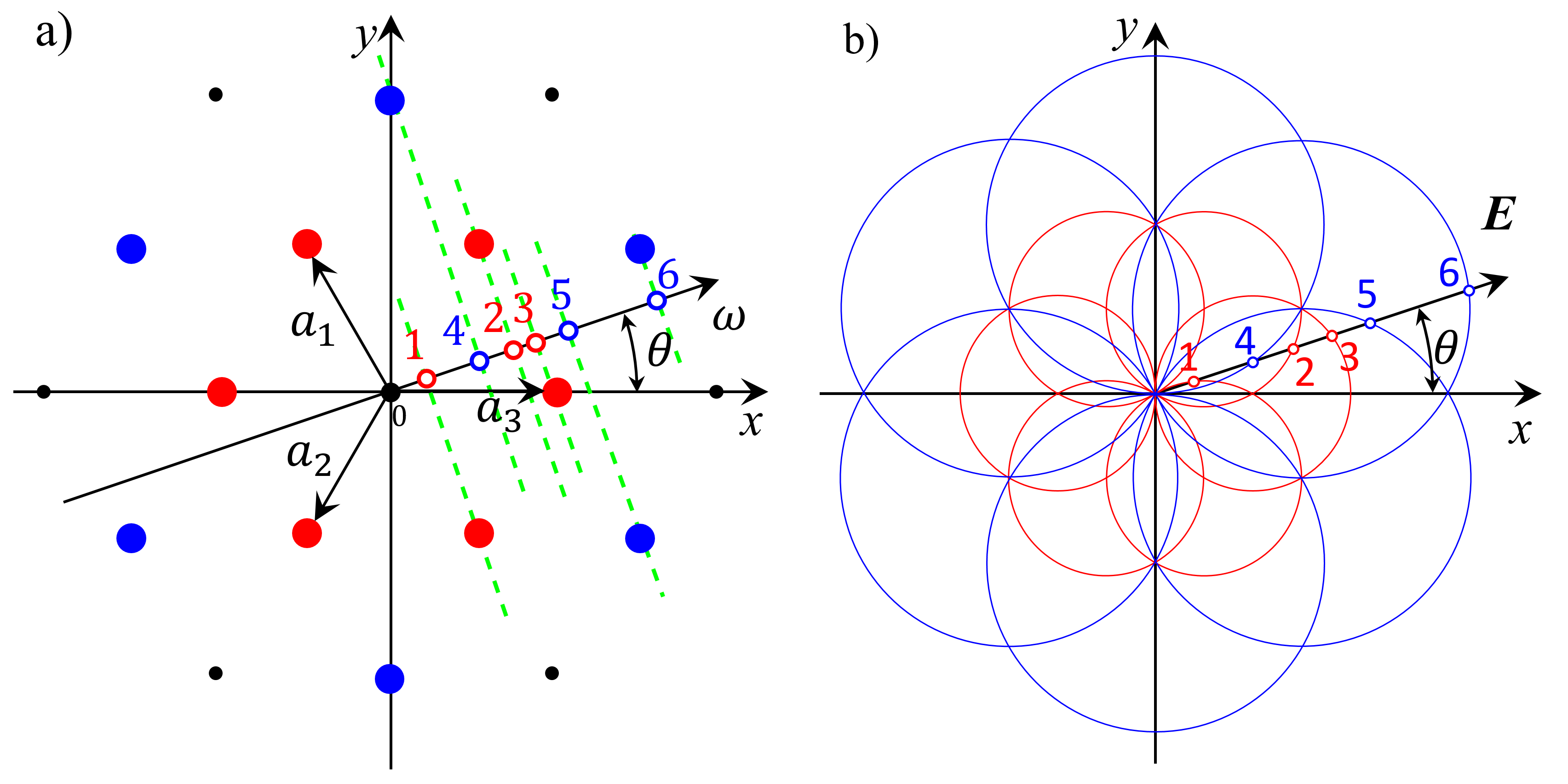}
\caption{a) 
Geometric construction of the frequency comb for Bloch oscillations, Eq.\eqref{eq:frequency_circles}, at a generic electric field orientation. % relative to the superlattice. 
%A depiction of the method to determine the discrete Bloch oscillation frequency values $\omega$ in a superlattice for a generic electric field orientation. 
%AC current spectrum for a generic dispersion. The black arrow is the direction of electric field $\vec E$. Solid circles are 2D $\delta$ functions obtained by Fourier transforming dispersion relation $\vec v(\vec k)$. Different colors corresponds to different magnitude of $\delta$-peaks, which in general decays exponentially as moving away from zero. The dashed green lines depict projecting the 2D $\delta$ peaks into 1D (see Eq.\eqref{eq:projection}). 
%The 2D${}\to{}$1D projection method 
%The discrete values of Bloch oscillation frequencies for asynchronous (free-carrier) oscillations in a 2D solid with trigonal symmetry.
Frequencies $\omega_l$ %, given by
%Illustrated is the geometric construction in 
%Eq.\eqref{eq:frequency_circles}, %  in a 2D solid with trigonal symmetry. %to determine discrete Bloch frequencies 
%illustrated 
%for a 2D solid with trigonal symmetry. 
%The 2D delta-peaks, organized in a 
are found by projecting the real-space Bravais lattice points (solid circles) %, are projected 
onto the 1D line parallel to $\vec E$ (black arrow) as indicated by dashed green lines. The shortest and next-shortest vectors are shown as red and blue dots. 
Hollow circles, found by projection, 
give the frequencies in Eq.\eqref{eq:frequency_circles}, where the emitted noise power $P(\omega)$ peaks. 
%see Eqs.\eqref{eq:NN_frequency}, \eqref{eq:NNN_frequency}. 
%peak positions in $P(\omega)$. % (see Eq.\eqref{eq:frequency_circles}). 
b) Visualization of the comb %the Bloch frequencies 
$\omega_l$ angle dependence 
vs. $\vec E$ orientation relative to the superlattice. 
}
\label{fig1_asynchronous}
%\vspace{-5mm}
\end{figure}

\begin{figure}[t]
\includegraphics[width=0.99\columnwidth]{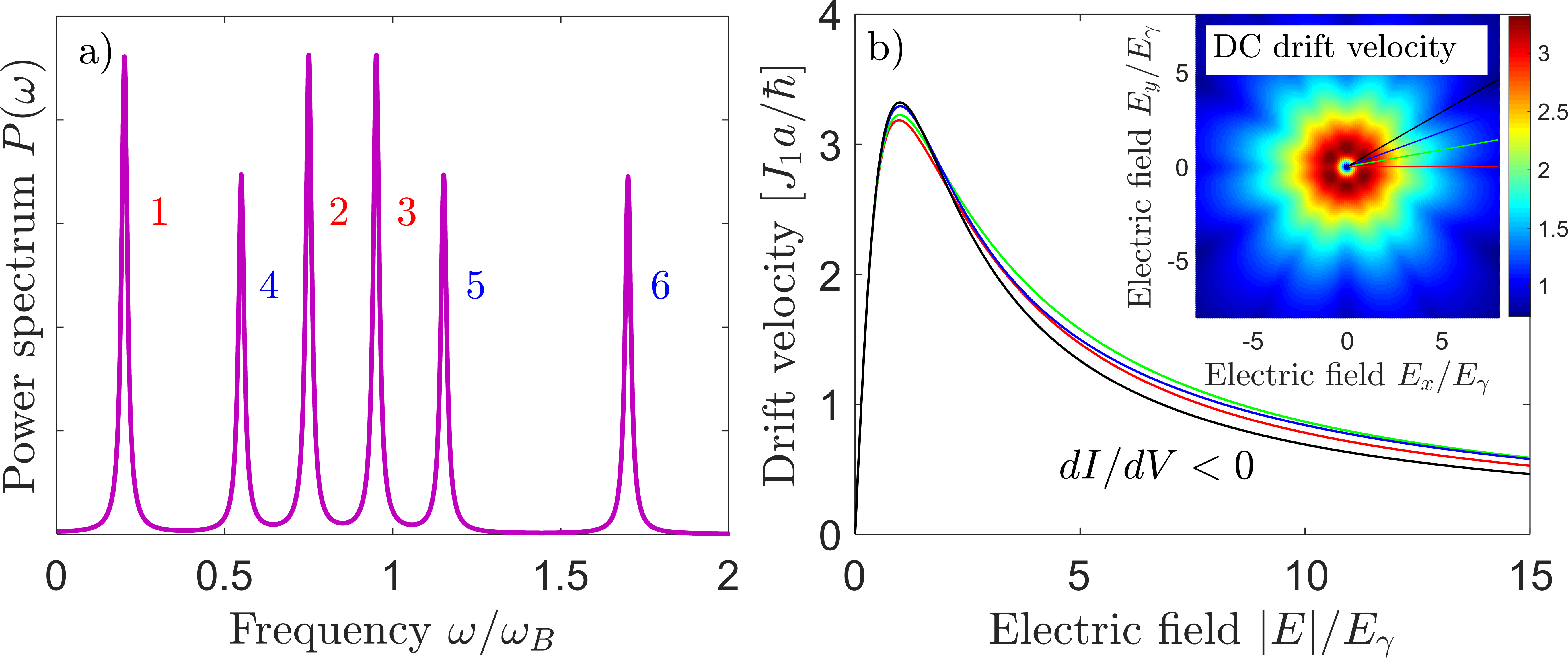} %Fig2_asynchronous_small.png}
\caption{
a) The comblike frequency spectrum of current fluctuations, Eq.\eqref{eq:spectrum_def}, 
%Testable signatures of asynchronous (free-carrier) Bloch oscillations. a) Ensemble-averaged AC current power spectrum, 
%Eq.\eqref{eq:spectrum_def},
%Eqs.\eqref{eq:spectrum_def}, %\eqref{eq:P(w)_lorentzian},}
%, and DC drift velocity, Eq.\eqref{eq:j_DC}. %Peaks in $P(\omega)$ occur 
%$P(\omega)$ 
%features 
consisting of finite-width resonances %narrow lines %peaks 
at the discrete frequency values $\omega_l$, Eq.\eqref{eq:frequency_circles}. 
% found %obtained 
%in (a) and (b); 
Frequency units are $\omega_{\rm B}=\frac{e}{\hbar}Ea$, the power spectrum $P(\omega)$ is in arbitrary units. The field orientation and %the red and blue 
labeling of different peaks match those in Fig.\ref{fig1_asynchronous}. 
b) The direct-current drift velocity, Eq.\eqref{eq:j_DC}. Shown is the full dependence (inset) and traces for several different field orientations. 
Bloch oscillations occur for field strength $E>E_{\gamma}=\hbar\gamma/ea$; negative differential conductivity $dI/dV<0$ %provides a clear signature 
is a hallmark of this regime.  
}
\label{fig2_asynchronous}
%\vspace{-5mm}
\end{figure}

%Prior to %presenting the analysis of 
%tackling 
Before discussing the synchronization problem we summarize the basic picture of the %asynchronous 
free-carrier Bloch oscillations in superlattices. %The dynamics 
In superlattices of dimension $D\ge 2$ %differs from %that in 
%the $D=1$ case in that 
different carriers can move at different angles relative to the applied field\cite{Rauh74,Gluck02,Dmitriev01,Dmitriev02,Kolovsky2013}. %Despite this difference, 
Nevertheless, the main properties of the one-dimensional Bloch oscillations persist. % in higher-dimensional bands. 
The Bloch frequencies remain discrete, taking values identical for all carriers in the system. %The only 
A new aspect is that different harmonics of the band dispersion produce %frequencies taking 
oscillations with several different discrete frequencies. These frequencies are in general incommensurate with one another, forming a comb-like spectrum pictured in Figs.\ref{fig1_asynchronous} and \ref{fig2_asynchronous}. 
%The frequency values can be adjusted by changing the field orientation, as illustrated in Fig.\ref{fig1_asynchronous}. 

The frequency comb %values %have a peculiar 
dependence on the electric field orientation %with respect to the superlattice 
is described by the geometric construction illustrated in Fig.\ref{fig1_asynchronous}. Namely, possible frequencies are given by the projections of different Bravais lattice vectors $\vec a_l=n_1\vec a_1^{(0)}+n_2\vec a_2^{(0)}$ on the applied field $\vec E$: 
\be\label{eq:frequency_circles}
\omega_l=\frac{e}{\hbar}\vec E\cdot \vec a_l=\frac{e}{\hbar}Ea_l\cos(\theta-\theta_l)
\ee
The dependence of the frequencies $\omega_l$ on the field $\vec E$ orientation and strength, 
as well as the tunability of moir\'e  superlattices by the twist angle, provide knobs that will facilitate achieving Bloch oscillations in moir\'e graphene.

This result can be illustrated %understood in very general terms 
by considering a general 
tight binding band on a monoatomic lattice,
\begin{align}
\label{eq:TBM_triangular}
\epsilon(\vec k) = \sum_{l=1,2...}-2J_l\cos(\vec k \cdot \vec a_l)
.
%& -2 J_1\sum_{j=1,2,3}\cos(\vec k \cdot \vec a_j)-2 J_2\sum_{j=1,2,3}\cos(\vec k \cdot \vec b_j),\\
%\vec a_j = & a(\cos \theta_j, \sin \theta_j), \quad \vec b_j=\sqrt{3}a(\cos \phi_j, \sin \phi_j)
\end{align}
%where $\vec a_l$ are vectors in the 
The Bravais lattice vectors $\vec a_l$ describe hopping between different pairs of lattice sites, either nearest-neighbor or non-nearest-neighbor. 
Bloch-oscillating free carriers obey quasiclassical equations of motion 
\be
\hbar \frac{d\vec k}{dt}=e\vec E,
\ee 
generating a linear time dependence $\vec k(t)=\frac{e}{\hbar}\vec E t+k_0$ with the linear part identical for all carriers and a carrier-specific initial value $k_0$. With this bandstructure and an electric field of a generic orientation, $\vec E=E(\cos\theta,\sin\theta)$, the frequencies at which the time-dependent velocity of the electrons $\vec v(t)=\frac1{\hbar}\nabla_k\epsilon(\vec k)|_{\vec k=\frac{e}{\hbar}\vec E t+k_0}$ will oscillate are given by $\vec a_l$ projected on $\vec E$, Eq.\eqref{eq:frequency_circles}. The resulting dependence of the frequencies $\omega_l$ on the orientation of $\vec E$ is described by families of circles pictured in Fig.\ref{fig1_asynchronous}.

Physically, discrete frequency values arise because electron trajectories sweep the (reduced) Brillouin zone (BZ) of a two-dimensional crystal in the direction set by the $\vec E$ vector. Every time an electron reaches zone boundary it umklapps to the opposite side and continues forward, winding around the BZ at different frequencies in different crystal axes directions. %At the same time, the average frequency 
In that, the time-averaged rate of winding around BZ along the direction of $\vec E$ is the same for all carriers. This leads, for a general field orientation, to a quasiperiodic dynamics characterized by two fundamental frequencies which depend only on the field $\vec E$ and lattice periodicity as described in Eq.\eqref{eq:frequency_circles}, wherein $\omega_l=n_1\omega_1+n_2\omega_2$ in agreement with the geometric construction in Fig.\ref{fig1_asynchronous}.

%\addQ{ for the shortest and next-shortest Bravais lattice vectors
%\begin{align}
%&\vec a_{j=1,2,3} =  a(\cos \theta_j, \sin \theta_j), \quad 
%\theta_j=(2\pi /3)j
%,
%\\
%&\vec a_4 = \vec a_1-\vec a_2, \quad
%\vec a_5=\vec a_2-\vec a_3, \quad
%\vec a_6=\vec a_3-\vec a_1
%.
%%, \quad
%%\vec b_j=\sqrt{3}a(\cos \phi_j, \sin \phi_j)
%\end{align}
%}

%\addLL{In a two-dimensional solid the frequency values depend only on the crystal lattice basis vectors $\vec a_1$, $\vec a_2$ and electric field orientation and strength, described by the geometric construction illustrated in Fig.\ref{fig1_asynchronous}. Possible values are combinations with integer coefficients of the projections of different Bravais lattice vectors on the applied field:
%\be\label{eq:frequency_circles}
%\omega = n_1 \omega_1 + n_2 \omega_2
%,\quad
%\omega_{1}={\textstyle\frac{e}{\hbar}}
%\vec E\cdot\vec a_{1}
%,\quad
%\omega_{2}={\textstyle\frac{e}{\hbar}}
%\vec E\cdot\vec a_{2}
%, 
%%n_{1,2} = 0,\pm1,\pm2...,
%\ee
%$n_{1,2} = 0,\pm1,\pm2...$ \addLL{(for a geometric derivation %of these relations 
%see Fig.\ref{fig1_asynchronous}, for an algebraic derivation see Eq.\eqref{eq:Greens_function} and accompanying discussion).} 
%The dependence of the frequencies $\omega_{1,2}$ on field orientation and strength, %which is illustrated in Fig.\ref{fig1}, 
%as well as the tunability of moir\'e  superlattices by the twist angle, % Furthermore, in moir\'e graphene the bandwidth is tunable by twist angle, providing an extra 
%provide %additional 
%the knobs that may help achieve Bloch oscillations in moir\'e graphene.}

In the presence of momentum-relaxing scattering the frequency spectrum %consists 
broadens into a sum of finite-width resonances centered at $\omega=\omega_l$. 
%The emitted power spectrum is proportional to 
%We evaluate %the harmonics of 
The quantity of interest is 
the autocorrelation function of current fluctuations $P(\omega) = \frac12\int_{-\infty}^{\infty} \langle\delta\vec j(t)\cdot \delta\vec j(t+\tau)\rangle e^{-i\omega \tau} d\tau$ which describes the spectrum of electric noise emitted by the system.
%which governs the emitted power spectrum. 
Simple analysis %, described in the Appendix, 
predicts a comb-like emitted power spectrum %given by a sum of Lorentzians: %carrier drift velocities
%\addQ{[connect to the above]}
\begin{align}\label{eq:spectrum_def}
P(\omega) %&= \frac12\int\limits_{-\infty}^{\infty} \langle\delta\vec j(t)\cdot \delta\vec j(t+\tau)\rangle e^{-i\omega \tau} d\tau
%\nonumber
%\\
%&
= \sum_l\frac{P_l}{(\omega-\omega_l)^2+\gamma^2}
\end{align}
(see \cite{Supplementary Information}). %Appendix). 
%LL where $\langle...\rangle$ denotes ensemble averaging 
%(see 
%LL Eq.\eqref{eq:P(w)_lorentzian} in 
%Appendix for derivation and the discussion of the validity of this result). 
The Bloch oscillation regime corresponds to non-overlapping resonances. 
Since the frequencies $\omega_l$ are proportional to the applied field $\vec E$ the oscillations appear when the field strength exceeds a threshold set by momentum-relaxing scattering, $E_\gamma=\gamma\hbar/ea$. At lower fields the resonances  merge into a broadband %continuum 
noise spectrum, indicating a suppression of the oscillations. 
% washing out the narrow-band noise expected at higher fields. 

In the Bloch oscillation regime the DC drift velocity exhibits negative differential conductivity $dI/dV<0$, a characteristic behavior that provides a clear signature of this regime. %A simple model 
A direct calculation \cite{Supplementary Information} %Appendix) 
predicts
\be\label{eq:j_DC}
\vec v_{\rm DC}=%\frac{e}{\hbar}
\sum_l\vec a_l \frac{2J_l f_{l}}{\hbar} \frac{\gamma  \omega_l}{\gamma^2+\omega_l^2}
,\quad
f_{l}=\sum_k f_0(k)e^{i\vec a_l\vec k}
,
\ee
% Eq.\eqref{eq:j_DC} predicts 
with $f_0(k)$ the steady-state momentum distribution. 
%Eq.\eqref{eq:j_DC} features a 
The dependence on the field $E$ is linear at small $E<E_\gamma$ and falls off as $1/E$ at large $E>E_\gamma$. Interestingly, current depends on the dimensionless quantity $E/E_\gamma$ in a way that is independent of the specific value of $\gamma$. This behavior is illustrated in Fig.\ref{fig2_asynchronous}(b). 
The drift velocity for electric fields in different directions is shown in the inset. 

%LL \addLL{The parameter values required to observe Bloch oscillations will be estimated below and shown to be realistic for moir\'e graphene superlattices.} \addLL{(write it)}

Next, we turn to the discussion of Bloch oscillations synchronized by coupling to an external oscillator mode:
%the great power in the sky}
%We begin with the Hamiltonian for 
%LL We consider the near-resonant case when the oscillator frequency is close to one of the free-carrier harmonics given in Eq.\eqref{eq:frequency_circles}. It is therefore sufficient to focus on the one-frequency Bloch oscillations described by the Hamiltonian  
%LL \addLL{(discuss implementation of the oscillator)}
\be\label{eq:H} 
H=\sum_{i}\left[\epsilon\left(\vec p_{i}\right)-e\vec E\vec x_{i}-\alpha Q x_i\right]+\frac{1}{2m}P^{2}+\frac{\omega_{0}^{2}m}{2}Q^{2}
.
%,\quad
%\addLL{a(t)=\alpha\int_{t'_i}^t Q(\tau)d\tau.}
\ee 
%LL \addLL{[justify 1D better]} 
Here $\epsilon(\vec p)$ is the band dispersion, $\vec p_{i}$ and $\vec x_{i}$ are the momenta and coordinates of the
%Bloch-oscillating 
electrons; $P$ and $Q$ are the momentum
and amplitude of the oscillator. 
The Bloch electron coupling to the oscillator and the external field is through potentials $U(\vec x_i)=-e\vec E\vec x_{i}-\alpha Q x_i$ seen by each of the electrons. 
In this approach we ignore the direct carrier-carrier interactions, treating electron dynamics in a free-particle approximation. 
Bloch oscillations are driven by the electric field $E$, the term $-\alpha Q x_i$ describes coupling of the electrons to the oscillator mode. 
In practice the oscillator can be realized as e.g. THz photonic or plasmonic resonators\cite{Savvidis2004,Ju2011,Yan2012a,Yan2012b,Tu2020,Ateshian2020}.  

%This change, as compared to our previous discussion, is introduced in order to avoid nonlocality in the Hamiltonian equations of motion.}
% $\phi(x)=-Ex$ and $a(t)=\alpha\int_{t'_i}^t Q(\tau)d\tau$, where $t'_i<t$ denote the time instants when the $i$-th electron state was last reset by scattering and the Hamiltonian dynamics described by Eq.\eqref{eq:H} had started.}

%We consider
Starting from the equations of motion originating from the Hamiltonian above, %in Eq.\eqref{eq:H}. 
we wish to integrate out the %electron 
carrier degrees of freedom and derive a closed-form  dynamics for the oscillator. For that purpose we solve equations of motion for the $i$-th electron beginning from the time $t'_i<t$ when its state was last reset by scattering and the Hamiltonian dynamics described by Eq.\eqref{eq:H} had started.

%LL \addLL{We employ a quasiclassical framework since in the regime of interest $eEa\ll W$ it provides a description of Bloch oscillations as well as the oscillator dynamics which is equivalent to the full quantum treatment. Conditions for quasiclassical dynamics: large oscillation amplitude $\gg a$, small frequency $\ll W$. (cite something? Supplement?)} %well suited both for describing the Bloch oscillations   

The full set of equations of motion for the electrons and the oscillator 
is
\begin{align}
\dot{\vec p}_{i} & =  -\frac{\partial H}{\partial \vec x_{i}}=e\vec E+\alpha Q(t)
,\quad 
\dot{\vec x}_{i}  =  \frac{\partial H}{\partial \vec p_{i}}=\frac{\partial\epsilon\left(\vec p_{i}\right)}{\partial \vec p_{i}}
\label{eq:El_vel_eq}
\\ \nonumber
\dot{P} & =  -m\omega_{0}^{2}Q+\sum_{i}\alpha x_i
,\quad %\frac{\partial\epsilon\left(p_{i}-a(t)\right)}{\partial Q}\\
\dot{Q}  =  P/m.
\end{align}
Eliminating $P(t)$ yields a second-order equation of motion for the oscillator mode $Q(t)$, driven by an external force given by a sum of contributions due to the electrons
\be
\ddot{Q}\left(t\right)+\omega_{0}^{2}Q\left(t\right)  = f(t),
\quad
f(t)=\frac{\alpha}{m}\sum_{i}x_i(t)
.
%\left.\frac{\partial\epsilon\left(\xi\right)}{\partial\xi}\right|_{\xi=\frac{a}{\hbar}p_{i}\left(t\right)-\alpha Q\left(t\right)}\nonumber \\
%\ddot{Q}+\omega_{0}^{2}Q & = & \frac{\alpha\hbar}{ma}\sum_{i}\dot{x}_{i}.
\label{eq:EoM_singleMode}
\ee
%$a$ is the lattice spacing. 
Importantly, the cumulative effect  due to the electrons, given by the quantity $f(t)$, %on the right hand side
gives rise to a ``memory effect'' in the oscillator dynamics. Each term in the sum $\sum_{i} x_i(t)$ is given by a solution of the equations of motion for $x_i(t)$ and $p_i(t)$, Eq.\eqref{eq:El_vel_eq}, %each 
initialized at an earlier random time $t'_i<t$. The oscillator dynamics $Q(t)$, $P(t)$ during the time intervals $t'_i<\tau<t$ affects the electron states $x_i(t)$, $p_i(t)$, giving rise to a back-action $f(t)=\frac{\alpha}{m}\sum_{i}x_i(t)$ with the dynamical memory originating from the dependence on $Q(\tau)$ and $P(\tau)$ at the earlier times $\tau<t$.
%, namely a dependence on $Q(\tau)$ and $P(\tau)$ at such earlier times. 

%As we show below, t
The feedback due to this memory effect 
%can result in 
enables synchronization of Bloch dynamics, resulting in a macroscopic oscillating current generated by Bloch-oscillating electrons. 
%such that Bloch-oscillating electrons generate a macroscopic oscillating current. 
To describe the instability we compute the backaction term linearized in $Q(t')$ (the analysis is lengthy but straightforward, see \cite{Supplementary Information}). %Appendix). 
%LL text moved to Appendix
Substituting the result in Eq.\eqref{eq:EoM_singleMode} gives a characteristic  equation for $\omega$ of the form
\begin{align}\label{eq:characteristic_eqn}
&\omega_0^2-\omega^2=\frac{i\lambda}{\omega}
\lp \frac{\gamma^2}{(\gamma^2+\omega_{\rm B}^2)(\gamma-i\omega)}
%+\frac{\gamma}{(\gamma+i\omega_{\rm B})(\gamma-i\omega)}
+\frac{\gamma}{(\omega+i\gamma)^2-\omega_{\rm B}^2}
\rp
, 
%\nonumber
%\\ 
%&\lambda = N \frac{\alpha^2 a v_0 }{m\hbar} =N\lp\frac{\alpha a}{\hbar}\rp^2 \frac{\Delta}{m}
%,
\end{align}
where we defined $\lambda = N \frac{\alpha^2 a v_0 }{m\hbar}$ with $N$ the total number of Bloch-oscillating electrons. 

The system becomes unstable when Eq.\eqref{eq:characteristic_eqn} admits solutions in the upper halfplane of complex $\omega$. Before exploring this %attempting to achieve 
instability we inspect, %briefly consider, 
as a sanity check, the regime of highly damped Bloch oscillations, $\gamma\gg \omega_{\rm B},\omega_0$. In this case, Eq.\eqref{eq:characteristic_eqn} reads
%the characteristic equation reads
$\omega_0^2-\omega^2=\frac{i\lambda}{\omega\gamma}$.
At large $\gamma$, the roots of this equation are close to $\pm\omega_0$. Writing $\omega=\pm\omega_0+\Delta\omega$, at leading order in $1/\gamma$ we find
$\Delta\omega=-\frac{i\lambda}{2\omega_0^2\gamma}$.
Negative imaginary part indicates that no instability arises in this regime, i.e. the driven system is stabilized by high damping.

%Next, we consider a 
A very different situation occurs at weak damping $\gamma\ll \omega_{\rm B},\omega_0$. The new behavior %in this case 
is simplest to understand close to the resonance between the oscillator and Bloch frequencies,  $\omega_0\approx \omega_{\rm B}$. %Focusing on the frequencies 
For $\omega$ values near the resonance, where the last term in  Eq.\eqref{eq:characteristic_eqn} dominates, we % over the preceding term. 
can ignore the first non-resonant term. %and obtain
This gives 
\be\label{eq:characteristic_with_gamma0}
(\omega_0^2-(\omega+i\gamma_0)^2)((\omega+i\gamma)^2-\omega_{\rm B}^2)=\frac{i\lambda\gamma}{\omega}
.
\ee
Here we added
%, for the sake of generality, 
the oscillator damping rate $\gamma_0$. % for the oscillator. % dynamics.
%LL \addLL{[discuss $\gamma_0$]} 
%Writing $\omega=\omega_0+\Delta\omega$ and 
Working near the resonance 
%, $\omega-\omega_0\ll \omega_0\approx \omega_{\rm B}$, 
and expanding in a small $\delta\omega =\omega-\omega_0\ll \omega_0\approx \omega_{\rm B}$ to obtain 
the complex frequency roots positioned near $\omega_0$, the characteristic equation becomes
\be\label{eq:small_gamma}
(\omega-\omega_0 +i\gamma_0) ( \omega +i\gamma -\omega_{\rm B})=-i\eta/4 %-\frac{i\lambda\gamma}{4\omega_0^3}
,\quad
\eta =\frac{\lambda\gamma}{\omega_0^3}
.
\ee
%where we defined $\Delta\omega_{\rm B}=\omega_{\rm B}-\omega_0$. 
The properties of Eq.\eqref{eq:small_gamma} are %simplest to understand 
are particularly straightforward 
when $\gamma_0=\gamma$. In this case, the roots are
%complex frequency roots positioned near $\omega_0$ are given by
\be
\omega_{1,2}=-i\gamma+\frac{\omega_{\rm B}+\omega_0 \pm\sqrt{\lp \omega_{\rm B}-\omega_0\rp^2 -i\eta %\frac{i\lambda\gamma}{\omega_0^3}
}}2
\ee
The system is stable if ${\rm Im}\,\omega_{1,2}<0$ and unstable otherwise. Using the identity
\be
{\rm Im}\lp \sqrt{x -i\eta %\frac{i\lambda\gamma}{\omega_0^3}
} \rp =-{\rm sgn}\,\eta \sqrt{\frac{\sqrt{x^2+\eta^2}-x}2}
\ee
with $x=(\omega_{\rm B}-\omega_0)^2$, 
%we can write 
the condition for the instability becomes
\be\label{eq:instability_criterion_0}
\eta^2 > %\lp \lp\Delta \omega_{\rm B}\rp^2+8\gamma^2\rp^2-\lp\Delta \omega_{\rm B}\rp^4=
\lp \lp\omega_{\rm B}-\omega_0\rp^2+4\gamma^2\rp 16\gamma^2
.
\ee
This criterion predicts the Bloch frequency %detuning from the resonance 
$\omega_{\rm B}$ and the coupling strength $\lambda$ values for which an instability towards a synchronized dynamics may occur, giving the phase diagram %pictured 
shown in Fig.\ref{fig1_oscillator}. 
%We see that
%In agreement with the discussion above, \addLL{(???)}
%As one might expect 
As expected on general grounds, the instability is easiest to achieve when Bloch oscillations are in resonance with the oscillator, $\omega_{\rm B}=\omega_0$. Tuning away from the resonance suppresses the instability. The instability signals the onset of a collective regime in which Bloch-oscillating electrons become synchronized through coupling to the oscillator mode. %behavior

A wider variety of collective regimes can be achieved by varying the oscillator damping $\gamma_0$. High and low damping values,  $\gamma_0\gg\gamma$ and $\gamma_0\ll\gamma$, favor synchronization and lasing, respectively. In both cases the instability towards collective dynamics can occur not only on the resonance $\omega_{\rm B}\approx\omega_0$ but also away from it in a relatively wide range of $E$ fields, % such that 
$\omega_{\rm B}<\omega_0$ for synchronization and $\omega_{\rm B}>\omega_0$ for lasing (see \cite{Supplementary Information}). We note %parenthetically 
that the lasing regime can also be understood in terms of a negative AC conductivity that enables gain of THz radiation\cite{Kroemer2000,Hyart2008,Hyart2009a,Hyart2009b}. 

An intriguing question for future work is the role of electron interactions. % between Bloch-oscillating carriers. %, in particular whether 
Several interesting regimes can be envisioned depending on the relation between carrier concentration and the localization radius of %Wannier-Stark 
WS states $r_0\sim J/eE$. At high carrier concentration, $n r_0^2\gg 1$, the interactions will act to dephase the oscillationsoscillations, producing an asynchronous Bloch-oscillating electron gas. To the contrary, at low carrier concentration, $n r_0^2\ll 1$,
%carrier separation is  greater than the oscillation amplitude, 
the interactions will tend to create a spatially ordered Wigner solid of localized Bloch-oscillating carriers. Ordering will stabilize oscillations and facilitate synchronization.

Another question of interest is the effect of thermal fluctuations and noise. While the electron temperature under a strong direct current is expected to be high, in the architecture considered above the temperature of an external oscillator is naturally decoupled from that of electrons. The oscillator will remain cold and provide a synchronizing feedback on the electron subsystem.

%Interestingly, reducing the oscillator damping to $\gamma_0\ll\gamma$ facilitates synchronization, favoring it in a wider range of parameters. In this case the instability towards synchronized dynamics can occur not only on the resonance $\omega_{\rm B}\approx\omega_0$ but also away from it in a wide range of $E$ fields such that $\omega_{\rm B}>\omega_0$ (see Supplementary Information). %Appendix).
%\cite{Hyart2008,Hyart2009a,Hyart2009b}.

In summary, 
%previous efforts to achieve collective Bloch oscillations in superlattice arrays of stacked semiconductor quantum wells came tantalizingly close to achieving this regime, however encountered obstacles due to electric field instabilities and noise. %towards electric field domain formation. 
%Here we propose a new system, the flat bands in moir\'e graphene, and argue that their unique %they have excellent 
the unique electronic properties of the flat bands in moir\'e graphene, such as the bandwidth considerably narrower than the optical phonon energy, the $\sim 10$nm-large superlattice periodicity and relatively high mobility, will facilitate observing the Bloch oscillations. The two-dimensional nature of the system offers additional benefits: 
the carriers, which are fully exposed, can be coupled to a nearby oscillator mode that will synchronize their movements 
to enable phase-coherent collective oscillations, 
a regime in which current-pumped synchronization and THz lasing can be realized and explored.
% and create realized   and therefore strongly . 
%A detailed analysis of the free-carrier Bloch oscillations synchronized by introducing an oscillator mode coupled to carrier displacements indicates feasibility of moir\'e-graphene-based Bloch oscillators. The instability achieved near the resonance between the Bloch frequency and the oscillator frequency is followed by a transition from asynchronous Bloch oscillations of free carriers to the synchronized collective-oscillation regime. The synchronized phase-coherent Bloch dynamics, driven by a DC current, has a potential to provide electromagnetic radiation sources with a stable and tunable frequency, realized at the submicron scale. The two-dimensional geometry of the moir\'e graphene systems is beneficial for overcoming the electric field domain formation encountered in previous work. The narrow bands in moir\'e superlattices therefore offer a unique platform %to realize %opportunity 
%in which Bloch oscillators driven by a DC current can be realized and explored. 

%\addQ{[talk about comb, broadband amplification]}

This work was supported by the Science and Technology Center for Integrated Quantum Materials, NSF Grant No. DMR-1231319; and Army Research Office Grant W911NF-18-1-0116 (L.L.). E.K. acknowledges financial support by the Research Travel Grant of the Karlsruhe House of Young Scientists (KHYS)

%LL Summary/discussion: Talk about experimental parameter values and the requirements that must be fulfilled in order to realize Bloch oscillations.

%Next, we consider the case $\gamma_0\ll\gamma$ and argue that the behavior is similar to that found above. The roots of Eq.\eqref{eq:small_gamma} are %this equation are
%\be
%\Delta\omega_{1,2}=\frac{\Delta \omega_{\rm B}-i\gamma \pm\sqrt{(\Delta \omega_{\rm B}-i\gamma)^2 -i\eta %\frac{i\lambda\gamma}{\omega_0^3}
%}}2
%\ee
%For large enough $\lambda$ values, which are controlled by the coupling strength $\alpha$ defined above, this equation predicts instability. This can be seen most easily by working on resonance $\Delta\omega_{\rm B}=0$, which gives % and ignoring damping on the left hand side. This gives
%, somewhat artificially, making  damping identical in both terms
%\be
%(\omega^2-\omega_0^2)^2=-\frac{i\lambda\gamma}{\omega}
%\Delta\omega_{1,2}=\frac{-i\gamma \pm\sqrt{-\gamma^2 -i\eta %\frac{i\lambda\gamma}{\omega_0^3}
%}}2.
%\ee
%For $\lambda/\omega_0^3$ greater than $\gamma$ the roots reside in both the upper and lower halfplanes
%Making the substitution $\omega=\pm\omega_0+\Delta\omega$ as above, yields an equation
%$
%\Delta\omega^2=\mp\frac{i\lambda\gamma}{\omega_0^3}
%$
%that has roots both in the upper and lower halfplanes, 
%\be
%\Delta\omega=\pm(1\mp i)\sqrt{\eta/2} %\frac{\lambda\gamma}{2\omega_0^3}}.
%\ee
%At the same time, for $\lambda/\omega_0^3$ smaller than $\gamma$ the roots are in the lower halfplane, i.e. the system is stable. \addLL{(Is it true?)}

\let\oldaddcontentsline\addcontentsline% Store \addcontentsline
\renewcommand{\addcontentsline}[3]{}
\let\addcontentsline\oldaddcontentsline

\clearpage

%\appendix
\renewcommand\thefigure{S\arabic{figure}}
\renewcommand{\theHfigure}{S\arabic{figure}}
\setcounter{figure}{0}
\setcounter{equation}{0}
\renewcommand\theequation{S\arabic{equation}}
\renewcommand{\theHequation}{S\arabic{equation}}
\setcounter{page}{1}
\newpage
%\onecolumngrid
\begin{center}
\noindent \textbf{Supplemental Material} % for ``Synchronizing Bloch-oscillating free carriers in moir\'e flat bands'' by A. Fahimniya, Z. Dong, E. I. Kiselev, and L. Levitov}
\end{center}
%\tableofcontents

%\begin{appendix}

%\centerline{\bf Supplementary Information} %Online Supplement for %``Synchronizing 
%``Bloch-oscillating free carriers in moir\'e flat bands'' by A. Fahimniya, Z. Dong, E. Kiselev and L. Levitov} %Appendix}
%\setcounter{equation}{0}
%\renewcommand{\theequation}{\thesection.\arabic{equation}}

\section{%Numerical calculation 
Phonon emission rates and carrier dephasing}

%In this section, we study the phonon emission of Bloch electrons. We will estimate the phonon emission rate in moire flat band systems, and show that an electric field can suppress phonon emission.

Here we consider carrier dephasing for Bloch-oscillating %electrons in a 
moir\'e  superlattices due to phonon emission. Since the width of moir\'e bands is considerably smaller than the optical phonon energies, the electron-phonon interactions are dominated by coupling to acoustic phonons. %As we will see the relevant phonon wavelengths are of the order of the superlattice period. Since the long-wavelength acoustic phonons are weakly coupled to electrons, their contribution to dephasing can be controled by increasing the superlattice periodicity. 
%As we will see, 
We show that the acoustic phonon emission rates can be tuned in a wide range by 
%varying the width of electron bands and abruptly quenched by applying relatively weak the in-plane electric fields
two independent knobs---the width of the moir\'e band, controlled by the twist angle, and the in-plane electric field. The bandwidth impacts phonon emission %rate by tuning
through the density of states; phonon emission is suppressed when the twist angle is tuned away from the magic flat-band value. The electric field suppresses the emission rate by %triggering a discretization in 
creating a discrete electron energy spectrum. As a result, phonon emission is suppressed as the field increases and the system enters the Bloch-oscillating state. Importantly, the threshold field for this suppression is relatively low, such that Bloch oscillations can be induced in the free-carrier regime, avoiding Wannier-Stark (WS) localization on a superlattice scale.
% \addQ{tuned efficiently by engineering the width of electron band and applying an electric field}.

%Yet, we will see that the effects of dephasing can be significant for narrow moire bands in which the transition rates are enhanced by a large density of electronic states. Optimizing the dephasing can therefore be achieved by choosing the twist angles away from the magic values in order to enlarge the bandwidth and reduce the density of states that governs the acoustic phonon emission. In that the band width must remain small enough in order to block the optical phonon emission pathways. Our estimates below show that this can be achieved by reducing the twist angle to the values somewhat below the magic values. 

We start by writing down the full Hamiltonian, which contains the free-particle parts for electrons and phonons, and an electron-phonon interaction term, here taken in the deformation-potential form:   %described by the  deformation-potential coupling:
\be \label{eq:H_total}
H = H_{\rm el} + H_{\rm ph} + H_{\rm el-ph}
\ee 
The electrons are described by a tight-binding model on %1D 
a two-dimensional superlattice with a linear potential due to electric field: %\addLL{[fix notation]} %The electric field is modeled as a stair-like potential:
\be \label{eq:H_e}
H_{\rm el} = -\sum_{\la \vec n \vec n' \ra} %\sum_{\vec a} 
Jc^\dagger_{\vec n} c_{\vec n'} %+{\rm h.c.} 
- \sum_{\vec n} ea\vec E\cdot \vec nc^\dagger_{\vec n} c_{\vec n}
\ee
where $a$ is the superlattice period, $\vec n=(n_x,n_y)$ with integer $n_x$ and $n_y$ are the discrete coordinates that label the superlattice sites. Here, for simplicity, we model the superlattice as a square lattice. The electric field is applied along a general direction, %\addLL{(a typo?)}
the quantities $c_{\vec n}^\dagger$ and $c_{\vec n}$ are the creation and annihilation operators describing carriers on the lattice. Fermions in 2D continuum are described by superpositions of % $c_{\vec n}$
different Wannier orbitals,
\be\label{eq:phi_W_c}
	\psi_{\vec r}=\sum_{\vec n}W(\vec r-\vec n a) c_{\vec n}
	,\quad 
	\psi^\dagger_{\vec r}=\sum_{\vec n}W^*_{\vec n}(\vec r-\vec n a) c_{\vec n}^\dagger
	,
\ee
where $W(\vec r-\vec n a)$ are Wannier orbitals centered at the superlattice nodes.  %the function $\phi_{\vec n}(\vec r)$ is the eigenstate of electron Hamiltonian Eq.\eqref{eq:H_e}. 
%This Hamiltonain has been studied since Ref.\cite{Wannier}, whereas the eigenstate wavefunction was first calculated in Ref.\cite{Katsura}%James,
%.
%\addLL{[Do we need these references here? They look to me as (mostly) trying to justify the tight-binding picture in a periodic solid.]}

Next we introduce %$\phi_{\vec n}(\vec r)$ to represent 
the eigenstates of the electron Hamiltonian Eq.\eqref{eq:H_e}, denoting them as $\phi_{\vec n}(\vec r)$. As always for the %Wannier-Stark 
WS ladder problem, the analysis is simplest in the momentum representation, $\phi_{\vec n}(\vec r)=\int d^2r e^{i\vec p\vec r} \phi_{\vec n}(\vec p)$. Indeed, in momentum representation the Schroedinger equation turns into a first-order ODE which can be solved explicitly. The states in the 2D continuum are then given by a convolution of the on-lattice states and Wannier orbitals, Eq.\eqref{eq:phi_W_c}. 

Accordingly, carrying out the analysis yields the momentum-space wavefunctions
%The momentum-space representation of these states 
$\phi_{\vec n}(\vec p)$ given by products of the WS ladder wavefunctions $\Psi(\vec n, \vec p)$ %\cite{Katsura}%James,
 and the Wannier-orbital %wavefunction 
formfactors $W(\vec p)=\int d^2r e^{i\vec p\vec r} W(\vec r)$: 
\be
\phi_{\vec n}(\vec p) =  \Psi (\vec n, \vec p)W(\vec p) \label{eq:eigenstate_wavefunction}
%\frac{ V }{\lp 2\pi \rp^2} \int_\infty d^2\vec p
\ee
Below, for simplicity, we use a Gaussian %function to model Wannier orbitals:
model for the quantities $W(\vec p)$, 
\be
W(\vec r) = %\pi^{-\frac{1}{2}}\xi^{-1}
\frac{1}{\sqrt{\pi} \xi}
e^{-\frac{r^2}{2\xi^2}},\quad W(\vec p) = \lp 2\pi\rp ^{\frac{1}{2}}\xi e^{-p^2\xi^2/2}.
\ee
where $\xi$ defines the Wannier orbital radius. %It is convenient to write $W(\vec p)$ as follows:
It will be convenient to factorize the Gaussian dependence as
$
W(\vec p) =  w(p_x)w(p_y)
$, 
where $\vec p = (p_x,p_y)$ and
\be
w(p_{j}) = (2\pi)^{\frac{1}{4}}\xi^{\frac{1}{2}} e^{-p_{j}^2\xi^2/2},\quad j=x,y .
\ee
%Both terms on the right-hand side of Eq.\eqref{eq:eigenstate_wavefunction} can be written in a separable form:
Importantly, the %Wannier-Stark 
WS ladder wavefunction $\Psi(\vec n,\vec p)$ %takes  the following 
can also be brought to a separable form for an electric field $\vec E = (E_x, E_y)$ applied in a generic incommensurate direction:
\be
\Psi (\vec n, \vec p) = \psi_x (n_x, p_x)\psi_y (n_y,p_y),
%\frac{1}{\sqrt{V}} 
%\int d^2r e^{-i\vec p \cdot \vec r}W(\vec r) = 
%\frac{1}{\sqrt{V}} 
%\pi^{\frac{1}{2}}\xi \exp(-p^2\xi^2/2) = w(p_x)w(p_y).
\ee
with the factors $\psi_x (n_x, p_x)$ and $\psi_y (n_y,p_y)$ given by 
%\begin{widetext}
\bea\label{eq:WS_wavefunction_x} 
&\psi_{j} (n_{j},p_{j}) = e^{-i F^{\lp i\rp}_{n_{j}}(p_{j})}, \quad j=x,y 
\\ \nonumber
& F^{\lp i\rp}_{n_j}(p_j) = \frac{2J}{eE_{j} a} \sin p_{j} a - n_{j}p_{j} a.
%&\psi_y (n,p) = e^{-i F^{\lp y\rp}_{n}(p)},\quad F^{\lp y\rp}_{n}(p) = \frac{2J}{eE_y a} \sin p a - np a,\label{eq:WS_wavefunction_y} \\
%
\label{eq:Wannier_function_xy}
\eea
This yields a separable representation for the full momentum-space wavefunctions in Eq.\eqref{eq:eigenstate_wavefunction}.

We note parenthetically that a more complicated treatment is required when an electric field is applied in a commensurate direction.  
In this case, instead of two-dimensional ladder, the WS problem yields a one-dimensional ladder, in which each level represents a one-dimensional band describing particle moving perpendicular to the electric field. Here, for simplicity, we focus on the case of the field applied in a generic incommensurate direction.

Next, we introduce phonons and electron phonon coupling. We model the acoustic phonons by the continuum %with the following
Hamiltonian %\addLL{(Do we need the $V$ factor?)}
\be
H_{\rm ph} =  \int \frac{d^2 q}{(2\pi)^2} \hbar \omega_{\vec q} a_{\vec q}^\dagger a_{\vec q}, \quad \omega_{\vec q} = sq
\ee
where %$\sum_{k\in BZ} ... =  L\int^{\pi/a}_{-\pi/a}\frac{dk}{2\pi}...$,
$s$ is the speed of sound, %$V$ is system volume,
and the momenta $\vec q$ form a continuum extending beyond the superlattice Brillouin zone. This model accounts for the presence of phonon modes with wavelengths that can be either shorter or greater than the superlattice periodicity $a$.

The electrons and phonons interact through the deformation potential coupling:
\bea\label{eq:H_el_ph_1}
&H_{\rm el-ph} = \int d^2 r D \vec \nabla \cdot \vec u(\vec r) \psi_{\vec r}^\dagger \psi_{\vec r}, 
% \quad
\\  \nonumber
&u(\vec r) = \sum_k \sqrt{\frac{2\hbar}{\rho_0 \omega_{\vec k}%V
}}\lb a_k(t)e^{i\vec k\vec r}+ a_k^\dagger(t) e^{-i\vec k\vec r}\rb,
\eea
where $\rho_0$ is the atomic mass density, %$L$ is the length of the system, 
$a_k(t) =a_k e^{-i\omega_{\vec k} t}$, %$x$ is the discrete coordinate that labels the superlattice sites, the discrete divergence $\Delta_x u_x = \frac{u_{x+a}-u_{x-a}}{2a}$, 
$\psi_{\vec r}^\dagger $ and $\psi_{\vec r}$ are the creation and annihilation of an electron at a continuum position $\vec r$ defined above. 

To proceed with the analysis we rewrite the continuum electron-phonon coupling, Eq.\eqref{eq:H_el_ph_1}, in the basis of eigenstates found above, Eq.\eqref{eq:eigenstate_wavefunction}. This gives
\be\label{eq:H_el_ph_2}
H_{\rm el-ph} = \sum_{\vec n,\vec n'} \tilde c_{\vec n}^\dagger \tilde c_{\vec n'}\la \vec n| D \vec \nabla \cdot \vec u|\vec n'\ra
,%\quad
\ee
where $\tilde c_{\vec n}$ and $\tilde c_{\vec n}^{\dagger}$ denote fermion operators for the eigenstates in Eq.\eqref{eq:eigenstate_wavefunction}, and $|\vec n\rangle$ is a short-hand notation for these states. Accordingly, the matrix element in Eq.\eqref{eq:H_el_ph_2} equals
\be
\la \vec n| D \vec \nabla \cdot \vec u|\vec n'\ra=\int d^2 r \phi^*_{\vec n}(\vec r)
D\nabla \cdot \vec u(\vec r)
\phi_{\vec n'}(\vec r)
.
\ee
%For an estimate, we will ignore the overlaps between different Wannier orbitals, focusing on the contributions with $\vec n'=\vec n$. 
%
Starting from the electron-phonon Hamiltonian Eq.\eqref{eq:H_el_ph_2} and Fermi's golden rule, we express %LL write
 the phonon emission rate by a carrier transitioning from a state $\left. |\vec n\ra$
  to a state $\left. |\vec n-\vec m\ra$  as
 %assuming that electron temperature is much higher than the lattice temperature. In this case,
\begin{widetext}
%LL as follows:
\be
\gamma = \frac{2\pi}{\hbar} \frac{2D^2\hbar }{\rho_0 s V} {\sum_{\vec m}}' \sum_{\vec q} |q| \left|\int d^2 \vec r e^{i\vec q \vec r}\overline \phi_{\vec n-\vec m}(\vec r)\phi_{\vec n}(\vec r)\right|^2  \delta(\hbar s|q| - ea \vec m\cdot \vec E )\label{eq:emission_rate_1}
\ee
\end{widetext}
where $\vec q$ and $\hbar s |q|$ are the phonon momenta and energies, $\sum_{\vec m}'$ is the summation over all Bravais lattice vectors $\vec m$ that satisfy the condition for phonon emission, $\vec m\cdot \vec E > 0$. Here we ignore phonon occupation numbers, assuming that electron temperature is much higher than the lattice temperature.

%\be
% \label{eq:Wannier_function}
%\ee
%where $W(\vec p)$ is written into separable form: $w(p) = %\pi^{\frac{1}{4}}\xi^{\frac{1}{2}} \exp(-p^2\xi^2/2)$.

Next, we evaluate the matrix elements in Eq.\eqref{eq:emission_rate_1}. In the general form given above the overlap integrals are pretty cumbersome. However the task of evaluating the overlap integrals can be simplified by employing an approximation of a small Wannier orbital radius, $\xi\ll a$. We start with
plugging Eq.\eqref{eq:eigenstate_wavefunction}-%\eqref{eq:WS_wavefunction_x}\eqref{eq:WS_wavefunction_y}
\eqref{eq:Wannier_function_xy} into Eq.\eqref{eq:emission_rate_1},
\begin{widetext}
\be\label{eq:emission_rate_2}
\gamma = \frac{4\pi}{\hbar} \frac{D^2}{\rho_0 s^2 } {\sum_{\vec m}}' \int  \frac{|q| dq_xdq_y}{(2\pi)^2} \left|  \sum_{p_x} \overline \psi_{0} \lp p_x+\frac{q_x}{2} \rp \psi_{m_x} \lp p_x-\frac{q_x}{2} \rp \overline{w} \lp p_x+\frac{q_x}{2} \rp w \lp p_x-\frac{q_x}{2} \rp \right|^2 \times \left|\sum_{p_y}...  
\right|^2 \delta(|q| - Q_{\vec m} )
\ee
where 
\be\label{eq:Q_m}
Q_{\vec m} = \frac{ea \vec m\cdot \vec E}{\hbar s}=\frac{e \pi \vec m\cdot \vec E}{\omega_*}
,\quad
\omega_*=\frac{\pi \hbar s}{a}
,
\ee
\end{widetext}
is the emitted phonon momentum, $\omega_*$ is the superlattice Debye's frequency, and the quantity $|\sum_{p_y}...|^2$ is identical to $|\sum_{p_x}...|^2$ up to a replacement $p_x,q_x,E_x,m_x\rightarrow p_y,q_y,E_y,m_y$.
	Now, we evaluate the term $|\sum_{p_x}...|^2$ in this expression. Plugging Eq.\eqref{eq:WS_wavefunction_x}-\eqref{eq:Wannier_function_xy} into Eq.\eqref{eq:emission_rate_2} yields
	\bea\label{eq:emission_rate_3}
	& \left|\sum_{p_x} ...\right|^2 %&= \left|  \sum_{p_x} \overline \psi_{0} \lp p_x+\frac{q_x}{2} \rp \psi_{m_x} \lp p_x-\frac{q_x}{2} \rp |w(p_x)|^2\right|^2 \exp(-q_x^2\xi^2/2)\\
	= \left|  \sum_{p_x} %e^{-i \lp F^{\lp x\rp}_{m_x}(p_x- q_x/2) - F^{\lp x\rp}_{0}(p_x+q_x/2)  \rp} 
	G(p_x) |w(p_x)|^2\right|^2 e^{-q_x^2\xi^2/2}
	\\ \nonumber
	&
	\approx \left|
	%\frac{a}{2\pi}\int_{-\pi/a}^{\pi/a} G(p_x)
	\la G\ra\sum_{p_x}|w(p_x)|^2\right|^2 e^{-q_x^2\xi^2/2}
	=|\la G\ra|^2 e^{-q_x^2\xi^2/2}
	\eea
	Here $G(p_x)$ denotes the function 
\be
	e^{i F^{\lp x\rp}_{0}(p_x+q_x/2) -i  F^{\lp x\rp}_{m_x}(p_x- q_x/2) }
\ee 
	which is periodic in $p_x$ with the period $2\pi/a$. We evaluate the quantity in Eq.\eqref{eq:emission_rate_3} using that the period of $G(p_x)$ is much smaller that the width of $|w(p_x)|^2=(2\pi)^{1/2}\xi e^{-\xi^2 p_x^2}$, namely $\pi/a\ll 1/\xi$. Accordingly, we replace $G(p_x)$ by its average value over the period and carried out integration over $p_x$ as $\sum_{p_x} |w(p_x)|^2=1$. 
	%This yields $|...|^2 \approx |\la G\ra|^2 e^{-q_x^2\xi^2/2}$
	Evaluating the average $\la G(p_x)\ra =\frac{a}{2\pi}\int_{-\pi/a}^{\pi/a} dp_x G(p_x)$ gives a Bessel function
%	\begin{widetext}
	\be
	\la G(p_x)\ra =%\frac{a}{2\pi}\int\limits_{-\pi/a}^{\pi/a} G(p_x) =
	 e^{-im_x q_x a/2} J_{m_x}\lp \frac{4J\sin\lp \frac{q_x a}{2}\rp }{eE_xa}\rp
	\ee 
	Applying the same approach to the integral over $p_y$ in Eq.\eqref{eq:emission_rate_2} yields a closed-form expression
	\begin{widetext}
\be
\gamma \sim  \frac{4\pi}{\hbar} \frac{D^2}{\rho_0 s^2 } {\sum_{\vec m}}' \int  \frac{q^2 dqd\theta }{(2\pi)^2}  \left|J_{m_x}\lp\frac{4J}{eE_xa}\sin\lp \frac{q a \cos\theta }{2}\rp\rp \right|^2\left|J_{m_y}\lp\frac{4J}{eE_ya}\sin\lp \frac{q a \sin\theta }{2}\rp\rp \right|^2 e^{-q^2\xi^2/2} \delta(|q| - Q_{\vec m} )
.
\label{eq:emission_rate_ultimate_form}
\ee
\end{widetext}
This expression, which was derived in the limit $a\gg\xi$, is reasonably accurate for the practically interesting parameter range  $a\gtrsim\xi$. 

\begin{figure}[b]
	\includegraphics[width=0.48\textwidth]{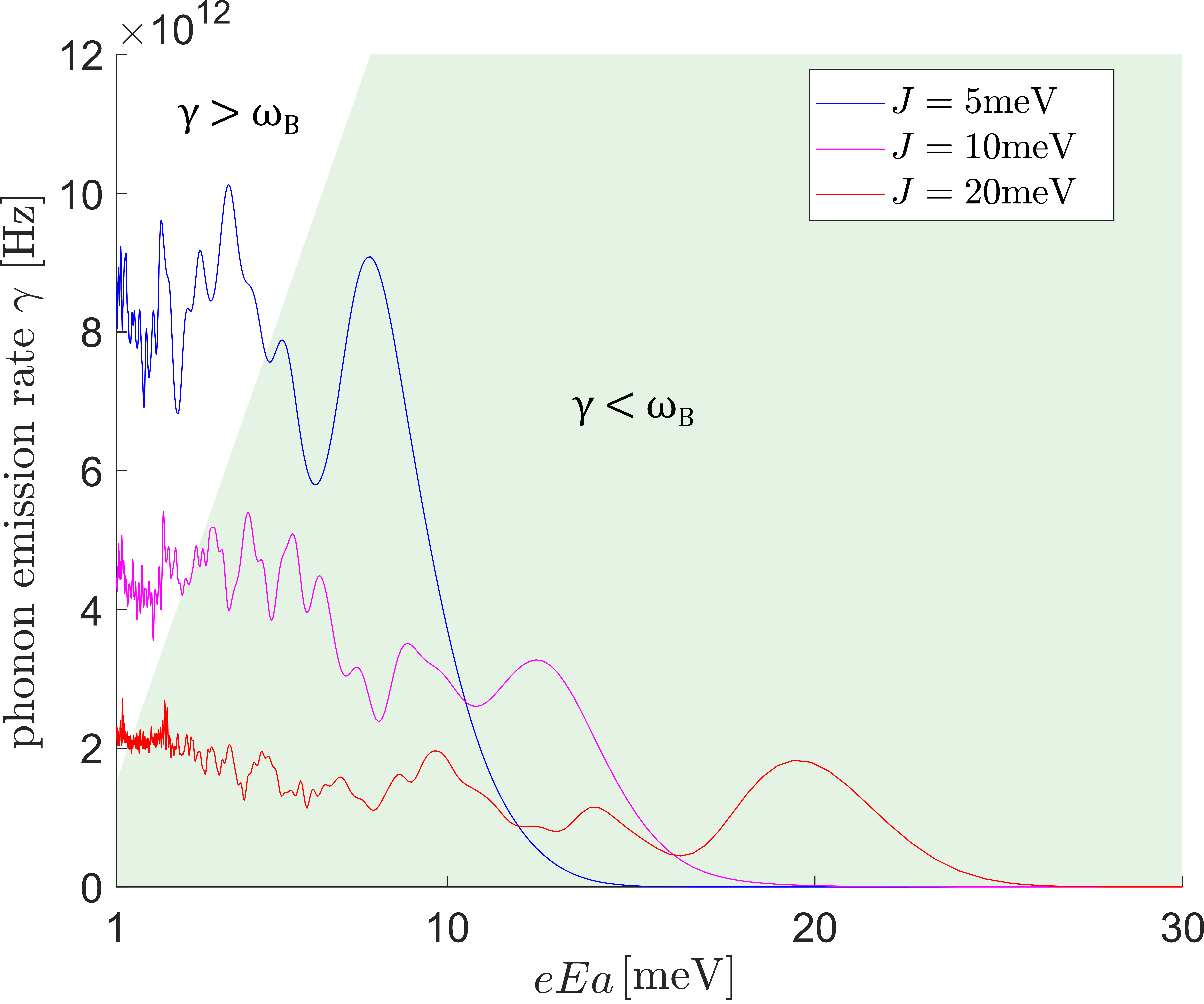}
	\centering
	\caption{The field dependence of phonon emission rate obtained from Eq.\eqref{eq:emission_rate_ultimate_form} for several different bandwidth values,
and typical moir\'e graphene parameter values given in the %main 
	text. In the green shaded region the Bloch oscillations are underdamped, $\gamma<\omega_B=eEa/\hbar$; in the white region the oscillations are overdamped  $\gamma>\omega_B$. %is safe from the 
%	dephasing due to phonon emission is weak. 
	The field orientation is incommensurate relative to the superlattice, such that $E_x/E_y=1.618$. The suppression of emission rate under increasing bandwidth and growing electric field is a generic behavior expected to remain valid for other incommensurate electric field orientations.
	%occur for both commensurate or incommensurate electric field orientations relative to the superlattice crystal axes.
	}\label{fig:emission_rate}
\end{figure}

The emission rate in Eq.\eqref{eq:emission_rate_ultimate_form} shows an interesting behavior as a function of system parameters. Crucially, it is sharply suppressed when either the bandwidth $J$ or the electric field $E$ increases.  These quantities can therefore serve as knobs to tune $\gamma$ and thereby control the Bloch-oscillating carrier dephasing.  The suppression of phonon emission in these two cases is governed by very different mechanisms. %by the two knobs are qualitatively different since they arises from distinct reasons.}
%This provides a unique opportunity   
The impact of the bandwidth on $\gamma$ can be understood in terms of the density of electronic states which control the emission rate, decreasing inversely with $J$. The dependence $\gamma$ vs. $E$ is fairly complicated due to the oscillatory character of the Bessel functions. The general trend, however, is simple to understand by noting that the energy spacings in the two-dimensional WS ladder grow as $E$ increases. As a result, the energies of different WS states are tuned out of resonance; this detuning suppresses phonon-mediated transitions. The suppression of phonon emission becomes exponential at $E$ much larger than the threshold value set by the maximal energy of phonons emitted through this process,  $eEa\gg\omega_{\rm max}\approx\omega_*\xi/a$. 

%The above argument explains the decrease of emission rate when electric field is applied in a commensurate direction. A more complicated case arises when electric field is in an incommensurate direction. Under this condition, it is always possible to find a lattice vector $\vec m = (m_x,m_y)$ such that $e\vec E \cdot \vec m a< \omega_{max}$. Does it means that the phonon emission will no longer get blocked in incommensurate field? Not really. We note that the Wannier-Stark wavefunction defines a characteristic lengthscale $\lambda_{WS} = \frac{J}{eEa}$, forbidding transitions in which electron hops by a distance larger than $\lambda_{WS}$. %With increasing $E$, there will be less $\vec m$ vectors that satisfy the condition $e\vec E \cdot \vec m a< \omega_{max}0$. 
%As a result, when $E$ is so large that the radius $\lambda_{WS}$ is smaller than the smallest $\vec m$ vector that satisfy the condition $e\vec E \cdot \vec m a< \omega_{max}$, the emission rate will decrease abruptly to zero. 
%\addLL{[I am not sure I understand the distinction made between $E$ applied in the commensurate and incommensurate directions. Let's discuss.]}

We illustrate the suppression of $\gamma$ in Fig.\ref{fig:emission_rate}, which shows 
%We calculate 
the emission rate obtained from Eq.\eqref{eq:emission_rate_ultimate_form} %The result is shown in Fig.\ref{fig:emission_rate}. Here, the 
for an electric field set to a generic direction. %, $E_x/E_y = 1.618$. 
Numerical values for other quantities are chosen to mimic a MATBG bandstructure: the superlattice period $a=10\rm{nm}$, the Wannier function radius  $\xi=0.5 a$. For these values, the superlattice Debye's frequency in Eq.\eqref{eq:Q_m} is  $\omega_{*} = 1\rm{meV}$. For el-ph coupling we use the graphene monolayer deformation potential 
%, speed of sound $s = \frac{\omega_{*}}{\pi/a}$, where $\omega_{*}$ is the superlattice Debye frequency, $\omega_{*} = 1\rm{meV}$ for MATBG. Deformation potential 
$D=20\rm{eV}$ and graphene mass density $\rho_0=7.6\times 10^{-8}\rm{g}/\rm{cm}^2$. 
%These momentum-space representation will be used later.

%The field dependence and bandwidth dependence of the phonon emission rate provide knobs to tune and suppress the dephasing of Bloch motion.
The above analysis, carried out for a square lattice tight-binding model, predicts %reveals a qualitative 
a behavior of phonon emission that we expect to remain qualitatively valid for other types of superlattices, in particular the moir\'e graphene superlattices. % in twisted bilayer graphene.
Namely, the large spatial periods of moir\'e superlattices and their abnormally narrow bandwidths limit phonon emission to the pathway dominated by acoustic phonons. 
We find, in particular, that the emission rate is quickly suppressed upon increasing the %superlattice 
bandwidth, see  Fig.\ref{fig:emission_rate}. Since  %moir\'e superlattices realized in 
the moir\'e graphene bandwidth is highly sensitive to the twist angle, becoming  small near the magic values, phonon emission can be suppressed by detuning the twist angle away from these values.

Likewise, the large superlattice periodicity results in a high sensitivity to %a strong dependence on 
the electric field. Our analysis predicts an abrupt quenching of phonon emission occurring already at moderate fields. The phonon emission rate features strong dependence on the bandwidth and field strength, these quantities can therefore serve as useful knobs allowing to realize and control Bloch oscillations.

%the simplest model. However, the qualitative picture it reveals---the field dependence and bandwidth dependence---%of the phonon emission rate,
%is generally applicable to more complicated systems, such as moir\'e superlattices. % like twisted bilayer graphene.
%Based on the analysis above, we conclude that the phonon emission rate is suppressed by increasing bandwidth, which is achievable in moir\'e superlattices by detuning the twist angle away from magic angle. Meanwhile, the large spatial period in moir\'e superlattices also allows the abrupt quenching of emission rate described in Fig.\ref{fig:emission_rate} to be realized in a moderate electric field. These two knobs overcome the major hurdle
%for observing coherent Bloch oscillations in moir\'e superlattices. 

%\begin{thebibliography}{99}
%\bibitem{Wannier}
%G. H. Wannier, Elements  of Solid State Theory, Cambridge University Press, Cambridge, England (1959).

%\bibitem{James}
%H. M. James, Electronic States in Perturbed Periodic Systems, Phys. Rev. 76, 1611 (1949)

%\bibitem{Katsura}
%S. Katsura, T. Hatta, A. Morita, On the Conception of the Energy Band in the Perturbed Periodic Potential, Progress of Theoretical Physics (1950)

%\end{thebibliography}

\section{The backaction on the oscillator due to Bloch-oscillating carriers and the role of oscillator damping}

Here we provide the details of the analysis of the backaction on the oscillator due to Bloch-oscillating carriers. We work with the equations of motion as given in Eqs.\eqref{eq:El_vel_eq}.
%LL moved here from the main text 
%LL Here we 
We average over the randomness in the starting times $t'_i$ ignoring the associated noise. This simple approach will be sufficient to %, which is crucial (???) for 
understand the synchronization effect. The role of randomness and noise will be discussed elsewhere. 

As a first step, we integrate Bloch dynamics of the $i$-th electron for times $t'_i<\tau<t$, which gives
\begin{align}
& \vec p_i(t) =e\vec E(t-t'_i)+\vec p_i(t'_i)+\alpha\int_{t'_i}^t Q(\tau')d\tau'
\\ \nonumber
& \vec x_i(t) =\vec x_i(t'_i)+\int_{t'_i}^t \vec v_i(\tau) d\tau
%,\quad 
%\frac1{\hbar}\nabla_k\epsilon(k)|_{k=p_i(\tau)/\hbar}
%v_0 \sin\left(\frac{a}{\hbar}p_i(\tau) \right)d\tau
%,\quad v_0 =\frac{a\Delta }{\hbar}
%LL \addAF{v_0 =\frac{a\Delta }{\hbar} }
,
\end{align}
%\addAF{[removed the extra $\frac{\alpha}{m}$ in the last equation and following propagations]}
where $\vec v_i(\tau)=\sum_l \frac{2J_l \vec a_l}{\hbar}\sin [\vec a_l\cdot\vec p(t)/\hbar]$. 
Averaging over the starting times $t'_i$ must be carried out using the survival probability %described by
obeying the Poisson statistics  $dp=dt\gamma e^{-\gamma(t-t'_i)}$.
%, drawing the initial momenta from the . 

It is instructive to first apply these relations to the free-carrier dynamics in the absence of coupling to the oscillator, $\alpha=0$. In this case different carriers are totally decoupled and thus not synchronized. The drift velocity can be found by averaging $\vec v_i(t)$ as
\begin{align}\nonumber
&\la \vec v_i(t)\ra
=
\sum_{\vec k'}\int_{-\infty}^t dt' \gamma e^{-\gamma (t-t')} \vec v_i(t,t')
%\sin \vec a_l\cdot \vec k(t)
\\ \label{eq:v(t)_ave}
&
=
\sum_{\vec k'}\sum_l \frac{J_l \vec a_l}{i\hbar}\gamma\lb \frac{e^{i\vec a_l\cdot \vec k'}}{\gamma- i \frac{e}{\hbar}\vec a_l\cdot \vec E} 
- \frac{e^{-i\vec a_l\cdot \vec k'}}{\gamma+ i \frac{e}{\hbar}\vec a_l\cdot \vec E}  \rb
,
\end{align}
where $\sum_{\vec k'}$ is a shorthand notation for averaging over the initial momentum distribution
$\int \frac{d^2k}{(2\pi)^2} f_0(\vec k')$ (here assumed to be steady-state). % (\addLL{comment on $f(k')$?}) 
The quantity $\vec v_i(t,t')$ under the integral over $t'$  is a sum of harmonics with frequencies $\omega_l$, arising from the carrier velocity time dependence
%we used 
\begin{align}
\vec v_i(t,t')=\sum_l \frac{2J_l \vec a_l}{\hbar}\sin \lb\vec a_l\cdot\lp \frac{e}{\hbar}\vec E(t-t')+\vec k'\rp\rb
.
\end{align}
Simplifying the result in Eq.\ref{eq:v(t)_ave} yields the drift velocity
\begin{align}\label{eq:v_dc}
&\vec v_{\rm DC}
=
\sum_{\vec k'}\sum_l \frac{2J_l \vec a_l}{\hbar}\cos(\vec a_l\cdot \vec k') \frac{\gamma \frac{e}{\hbar}\vec a_l\cdot \vec E}{\gamma^2+ (\frac{e}{\hbar}\vec a_l\cdot \vec E)^2}. 
\end{align}
Given by a sum of the terms $\frac{\gamma\omega_l}{\gamma^2+\omega_l^2}$, 
the dependence $v_{\rm DC}$ vs. $E$ is nonmonotonic, growing linearly at $E\lesssim E_\gamma=\gamma\hbar/ea$ and decreasing at $E\gtrsim E_\gamma$; at weak fields it matches the Drude theory prediction. 
%: at weak fields it matches the Drude theory prediction, with $v_{\rm DC}$ growing linearly with $E$; at $E\gtrsim E_\gamma=\gamma\hbar/ea$ the drift velocity $v_{\rm DC}$ drops, as illustrated in Fig.\ref{fig2_asynchronous}. 
The negative differential conductivity $dI/dV<0$ is a testable signature of the Bloch-oscillation regime. 

%$v_{\rm DC}$ grows linearly vs. $E$ at weak fields and matches the Drude theory prediction result, and a decreasing

The spectrum of current fluctuations, Eq.\eqref{eq:spectrum_def}, can be obtained in a similar manner. The velocity time dependence $\vec v_i(t,t')$ is a sum of harmonics with frequencies $\omega=\omega_l$; each harmonic producing %the comb of 
a resonance broadened by the damping rate $\gamma$. Indeed, evaluating the Fourier components and averaging over the initial times gives
\begin{align}
&\int_{-\infty}^t dt' \gamma e^{-\gamma (t-t')} \vec v_i(t,t') e^{-i\omega (t-t')}
\\
&
=\sum_l \frac{J_l \vec a_l}{i\hbar}\gamma\lb \frac{e^{i\vec a_l\cdot \vec k'}}{\gamma- i (\omega+\omega_l)} 
- \frac{e^{-i\vec a_l\cdot \vec k'}}{\gamma- i (\omega-\omega_l)}  \rb
.
\end{align}
Taking squares of the absolute values yields a fairly cumbersome expression for the noise spectrum. In the small-$\gamma$ limit, achieved at %which simplifies in the limit 
$E\gtrsim E_\gamma$, it represents a comb of sharp Lorentzians %given in Eq.\eqref{eq:spectrum_def} 
plus a background part, see Eq.\eqref{eq:spectrum_def} and Fig.\ref{fig2_asynchronous}. 

Next, we reinstate the coupling to the oscillator and proceed with the analysis of  synchronization. For conciseness, we focus on a resonance approximation valid near one of the resonances $\omega=\omega_l$ in Eq.\eqref{eq:spectrum_def}, at $\omega_{\rm B}\gg\gamma$. In what follows, without loss of generality, we take $\vec E$ to be parallel to $\vec a_l$, and denote $\omega_l$ and $\vec a_l$ as $\omega_{\rm B}$ and $a$, respectively. Generalizing to the large-$\gamma$ case and other field orientations will be straightforward. The special cases of field orientation such that $\vec E\cdot \vec a_l\approx \vec E\cdot \vec a_{l'}$, when two resonances can be excited simultaneously, will be discussed elsewhere.

%To proceed, we will assume a simple band
%structure of the form
%\[
%\epsilon\left(p\right)=-\Delta\cos\left(\frac{a}{\hbar}p\right)
%,
%\]
%where $a$ is the lattice spacing. Integrating 
The back-action of the carriers on the oscillator, given by the sum of carrier displacements $f(t)=\frac{\alpha}{m}\sum_i x_i(t)$ in Eq.\ref{eq:EoM_singleMode} averaged %over carrier displacements $x_i(t)$ 
%Averaging 
over the starting times $t'_i$ with the Poissonian survival probability %described by
%obeying the Poisson statistics  
$dp=dt \gamma e^{-\gamma(t-t'_i)}$, equals
\begin{align}
&\la x_i(t)\ra =  \la x_i(t')\ra + \int\limits_{-\infty}^t dt' \gamma  e^{-\gamma(t-t')}\int\limits_{t'}^t  d\tau v_0 \sin 
\frac{a p_i(\tau)}{\hbar}
\nonumber
\\
&= \!\int\limits_{-\infty}^t \!\!dt' \gamma  e^{-\gamma(t-t')}\int\limits_{t'}^t  d\tau v_0  \sin\lp \phi(\tau)\rp
%\lp \omega_{\rm B}(\tau-t')+\frac{\alpha a}{\hbar}\int\limits_{t'}^\tau Q(\tau')d\tau'\rp 
%,\quad \addAF{v_0 =\frac{a\Delta }{\hbar} }
,\ \ v_0 =\frac{2aJ_l }{\hbar}
, 
\end{align}
where %$\omega_{\rm B}=eEa/\hbar$ is the Bloch frequency, and 
we denote % \addAF{$v_0 =\frac{a\Delta }{\hbar}$} and 
$\phi(\tau)=\omega_{\rm B}(\tau-t')+\frac{\alpha a}{\hbar}\int_{t'}^\tau Q(\tau')d\tau'$. 
%Here we have 
In what follows we drop the starting displacement term $\la x_i(t')\ra$, assuming that it vanishes under averaging as expected for a spatially uniform distribution.
%LL (to be revisited later). 

The single mode dynamics is now described by Eq.\eqref{eq:EoM_singleMode} with the right-hand side replaced with a back-action memory function $\frac{\alpha}{m}N \la x_i(t)\ra$, where $N$ is the number of Bloch electrons. We will consider the dynamics at lowest nonvanishing order in $Q(t)$, assuming the latter to be small. First, setting $Q(\tau')=0$ and integrating over $\tau$, we find 
$ %\begin{align}
\la x_i^{(0)}(t)\ra 
%&=\int_{-\infty}^t dt' \gamma e^{-\gamma(t-t')}\frac{v_0 }{\omega_{\rm B}}\lb 1-\cos\omega_{\rm B}(t-t')\rb
%\nonumber
%\\
=\frac{v_0 \omega_{\rm B}}{\gamma^2+\omega_{\rm B}^2}
$,
%\end{align}
a constant displacement that gives a time independent contribution to $f(t)$ in Eq.\eqref{eq:EoM_singleMode}, which can be %accounted 
compensated for by %redefining 
shifting the oscillator equilibrium. 
%a constant displcement of the oscillator 
Next, at first order in $Q(t)$, we Taylor-expand the sine term to obtain
\begin{widetext}
\be\label{eq:x1} 
\la x_i^{(1)}(t)\ra = 
 \int_{-\infty}^t dt' \gamma  e^{-\gamma(t-t')} \lp \int_{t'}^t  d\tau v_0  \cos\lp \omega_{\rm B}(\tau-t')\rp \lb\frac{\alpha a}{\hbar}\int_{t'}^\tau Q(\tau')d\tau' \rb\rp
 .
 \ee
Plugging in a harmonic dependence $Q(t)=Q_0e^{-i\omega t}$, we evaluate the integrals over $\tau'$ and $\tau$ as
\begin{align}
&\int_{t'}^t  d\tau v_0  \cos\lp \omega_{\rm B}(\tau-t')\rp \lb\frac{\alpha a}{\hbar}\int_{t'}^\tau Q(\tau')d\tau' \rb=\int_{t'}^t  d\tau v_0  \cos\lp \omega_{\rm B}(\tau-t')\rp \lb\frac{i\alpha a}{\hbar \omega}Q_0\lp e^{-i\omega \tau}- e^{-i\omega t'}\rp \rb
\nonumber
\\
&= \frac{i\alpha a v_0 }{\hbar \omega}Q_0
\lp e^{-i\omega t} \frac{e^{i\omega_{\rm B}(t-t')}-e^{i\omega (t-t')}}{2i(\omega_{\rm B}-\omega)}+e^{-i\omega t}\frac{e^{-i\omega_{\rm B}(t-t')}-e^{i\omega (t-t')}}{-2i(\omega_{\rm B}+\omega)}-e^{-i\omega t'}\frac{\sin\omega_{\rm B}(t-t')}{\omega_{\rm B}}\rp
.
\end{align}
%EK: There is a factor 1/2 missing in front of the first two terms in brackets in Eq. (18). It stems from the expansion of cos into exp in the first line. However, these terms are off-resonant and do not alter Eq. (25).
Integration over $t'<t$ in Eq.\eqref{eq:x1} can now be carried out with the help of the identity
\[
\int_{-\infty}^t dt' \gamma  e^{-\gamma(t-t')} e^{-i\Omega (t-t')}=\frac{\gamma}{\gamma+i\Omega}
,
\]
giving %the result
\begin{align}
&\la x_i^{(1)}(t)\ra = \frac{i\alpha a v_0 }{\hbar \omega}Q_0 e^{-i\omega t}
\lp \frac{
\frac{\gamma}{\gamma-i\omega_{\rm B}}-\frac{\gamma}{\gamma-i\omega}}{2i(\omega_{\rm B}-\omega)}
+\frac{\frac{\gamma}{\gamma+i\omega_{\rm B}}
-\frac{\gamma}{\gamma-i\omega}}{-2i(\omega_{\rm B}+\omega)}
%\frac{\frac{\gamma}{\gamma+i\omega_{\rm B}}
-\frac{\frac{\gamma}{\gamma-i(\omega+\omega_{\rm B})}-\frac{\gamma}{\gamma-i(\omega-\omega_{\rm B})}}{2i\omega_{\rm B}}\rp
%-e^{-i\omega t'}\frac{\sin\omega_{\rm B}(t-t')}{\omega_{\rm B}}\rp
\nonumber
\\
&
= \frac{i\alpha a v_0 }{\hbar \omega}Q_0 e^{-i\omega t}
\lp \frac{\gamma}{2(\gamma-i\omega_{\rm B})(\gamma-i\omega)}
+\frac{\gamma}{2(\gamma+i\omega_{\rm B})(\gamma-i\omega)}
-\frac{\gamma}{(\gamma-i(\omega+\omega_{\rm B}))(\gamma-i(\omega-\omega_{\rm B}))}
\rp
\\
\nonumber
&
= \frac{i\alpha a v_0 }{\hbar \omega}Q_0 e^{-i\omega t}
\lp \frac{\gamma^2}{(\gamma^2+\omega_{\rm B}^2)(\gamma-i\omega)}
%+\frac{\gamma}{(\gamma+i\omega_{\rm B})(\gamma-i\omega)}
+\frac{\gamma}{(\omega+i\gamma)^2-\omega_{\rm B}^2}
\rp
.
\end{align}
\end{widetext}
Substituting this result in Eq.\eqref{eq:EoM_singleMode} gives a characteristic  equation for $\omega$ of the form given in Eq.\eqref{eq:characteristic_eqn}. The instability criterion and the phase diagram for the oscillator damping equal to that of Bloch-oscillating carriers is discussed in the main text (see Fig.\ref{fig1_oscillator} and accompanying discussion). 

\begin{figure}[t]
\includegraphics[width=0.99\columnwidth]{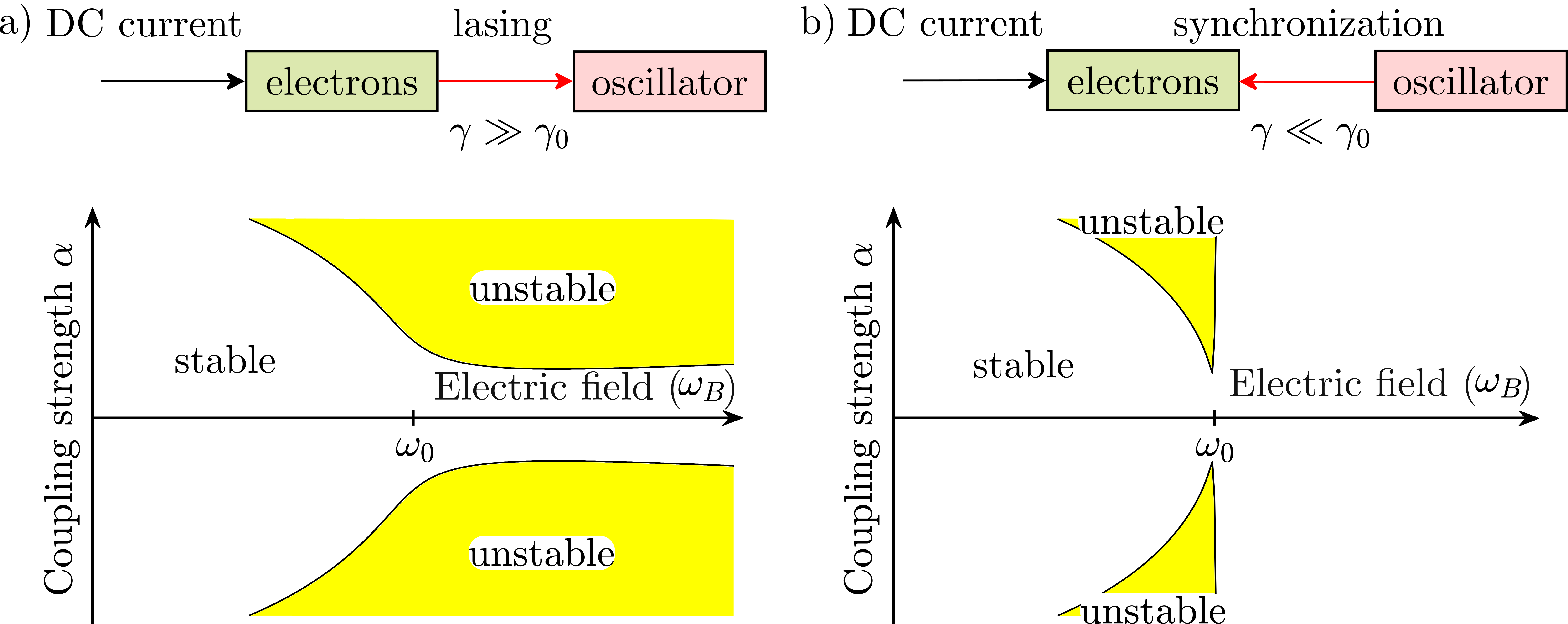}
\caption{The lasing and synchronization regimes.
a) Lasing ($\gamma \gg \gamma_0$). In this case, the oscillator is weakly damped and serves as the main reservoir of the energy. Energy of the electrons is more easily pumped to the oscillator when $\omega_{\rm B} > \omega_0$. Shown is the phase diagram for $\gamma = 100 \gamma_0$.
b) Synchronization ($\gamma \ll \gamma_0$). In this case, the oscillator is strongly damped and the electrons serve as the main reservoir of the energy. Energy of the oscillator is more easily pumped to the electrons when $\omega_{\rm B} < \omega_0$. Shown is the phase diagram for $100 \gamma = \gamma_0$.
The instability criterion is a sign change of the imaginary parts of the roots of Eq.~\eqref{eq:characteristic_with_gamma0}, which is negative in the stable regime and becomes positive in the unstable regime.
}
\label{fig2_oscillator}
\end{figure}

It is %straightforward 
instructive to extend this analysis to the more general case of unequal damping rates for the oscillator and electrons, $\gamma_0\ne\gamma$. After some algebra we arrive at the instability criterion 
\begin{align}\nonumber
\lp \eta+2(\gamma-\gamma_0)(\omega_{\rm B}-\omega_0)\rp^2 > %\lp \lp\Delta \omega_{\rm B}\rp^2+8\gamma^2\rp^2-\lp\Delta \omega_{\rm B}\rp^4=
&\lp \lp\omega_{\rm B}-\omega_0\rp^2+4\gamma\gamma_0\rp 
\\ 
& \times 4(\gamma+\gamma_0)^2
.
\label{eq:instability_criterion_2}
\end{align}
%resembling that in Eq.\eqref{eq:instability_criterion_0}. 
%As in Eq.\eqref{eq:instability_criterion_0}, the instability is easiest to reach on resonance $\omega_{\rm B}=\omega_0$.\addAF{[from the solution in Fig.~\ref{fig2_oscillator}, this is not the case. The additional $2(\gamma-\gamma_0)(\omega_{\rm B}-\omega_0)$ term on the left-hand side makes the easiest $\omega_{\rm B}$ for the instability shift.]} 
A new interesting behavior found for $\gamma_0\ne\gamma$ is an asymmetry between $\omega_{\rm B}$ blue-shifted and red-shifted away from $\omega_0$, with the instability threshold lower for $\omega_{\rm B}>\omega_0$ and higher for $\omega_{\rm B}<\omega_0$ when $\gamma_0<\gamma$, and vice versa when $\gamma_0>\gamma$, as illustrated in Fig.~\ref{fig2_oscillator}. The asymmetry is particularly striking in the limit $\gamma_0/\gamma\to 0$: for $\omega_{\rm B}>\omega_0$ the instability occurs at the coupling values $\eta$ much smaller than those in Eq.\eqref{eq:instability_criterion_0}, whereas for $\omega_{\rm B}<\omega_0$ the instability threshold remains on the same order as in Eq.\eqref{eq:instability_criterion_0}. Furthermore, perhaps somewhat counterintuitively, for $\gamma_0/\gamma\to 0$ the lowest value of coupling at which the instability sets in occurs far away from the resonance $\omega_{\rm B}=\omega_0$. 

The origin of this asymmetry is closely related to the mechanism that enables the synchronized behavior. %can be understood as follows. 
When the oscillator is undamped, synchronization arises due to the electrons pumping energy into the oscillator mode; subsequently, when this energy is passed back to electrons, they become synchronized with the oscillator, and with each other.
%the oscillator, in turn, stores this energy and passes it back to electrons. 
However, at a weak coupling $\eta$, the energy transfer from the Bloch-oscillating electrons into the oscillator is possible only if $\hbar\omega_{\rm B}>\hbar\omega_0$, indicating that the instability is easier to reach  for $\omega_{\rm B}$ values blue-shifted from $\omega_0$.

%than for the red-shifted $\omega_{\rm B}$ values. 

The above argument also suggests a reversal in the asymmetry when Bloch oscillations are weakly damped compared to the oscillator damping, $\gamma\ll\gamma_0$. Indeed, in this case it is the electron subsystem that serves as the main reservoir for energy storage, whereas the role of the oscillator mode is merely to lock the phases of different Bloch-oscillating carriers. Pumping energy into the collective mode now requires $\hbar \omega_{\rm B}<\hbar\omega_0$. We therefore expect that in this limit the instability will occur at lower $\eta$ values % more easily reachable 
for $\omega_{\rm B}$ red-shifted from $\omega_0$. This is exactly what Eq.\eqref{eq:instability_criterion_2} predicts (see Fig.\ref{fig2_oscillator}).

%LL \addLL{Discuss factors that limit the growth due to instability towards a synchronized behavior, possible collective mode frequency values, what else?}

%LL elusive  AC/DC  THz

%LL Next steps: 
%let's start with Eq.\eqref{eq:characteristic_eqn} and 
%LL Let's check the phase diagram. Make a plot, and discuss factors that will limit the exponential  growth of the unstable mode.  

%\end{appendix}

\begin{thebibliography}{99}
\bibitem{ashcroft_mermin} N. W. Ashcroft and N. D. Mermin, Solid State Physics (Saunders, Philadelphia, 1976).
%chambers} R. G. Chambers, Proc. Phys, Soc. (London) A6S, 458 (1952).
\bibitem{pippard} A. B. Pippard, The Dynamics of Conduction Electrons (Gordon and Breach Science Publishers, Inc., New York, 1965).
\bibitem{Esaki1970}
L. Esaki and R. Tsu, Superlattice and Negative Differential Conductivity in Semiconductors,
IBM J. Res. Dev., {\bf 14}, 61-65 (1970).

\bibitem{Ktitorov1972}
S. A. Ktitorov, G. S. Simin, and V. Y. Sindalovskii, 
Bragg reflections and the high-frequency conductivity of an electronic solid-state plasma, 
Fizika Tverdogo Tela, {\bf 13}, 2230-2233 (1971) [Soviet Physics - Solid State {\bf 13}, 1872-1874 (1972)]

\bibitem{Kroemer2000}
H. Kroemer, On the nature of the negative-conductivity resonance in a superlattice Bloch oscillator, \href{https://arxiv.org/abs/cond-mat/0007482}{arXiv:cond-mat/0007482} (2000).

\bibitem{Savvidis2004}
P. G. Savvidis, B. Kolasa, G. Lee, and S. J. Allen, 
Resonant Crossover of Terahertz Loss to the Gain of a Bloch oscillating InAs/AlSb Superlattice, Phys. Rev. Lett. {\bf 92}, 196802 (2004).

\bibitem{Sibille1990}
A. Sibille, J. F. Palmier, H. Wang, and F. Mollot, 
Observation of Esaki-Tsu negative differential velocity in GaAs/AlAs superlattices,
Phys. Rev. Lett. {\bf 64}, 52 (1990).

\bibitem{Feldmann1992}
J. Feldmann, K. Leo, J. Shah, D. A. B. Miller, J. E. Cunningham, 
T. Meier, G. von Plessen, A. Schulze, P. Thomas, and S. Schmitt-Rink,
Optical investigation of Bloch oscillations in a semiconductor superlattice,
Phys. Rev. B {\bf 46}, 7252 (1992).

\bibitem{Waschke1993} 
C. Waschke, H. G. Roskos, R. Schwedler, K. Leo, H. Kurz, and K. Kohler,
Coherent submillimeter-wave emission from Bloch oscillations in a semiconductor superlattice, Phys. Rev. Lett. {\bf 70}, 3319 (1993).

\bibitem{Rauh74}  A. Rauh and  G. H. Wannier, Theory of stark ladders in the optical absorption of solids, Solid State Commun. {\bf 15}, 1239 (1974).

\bibitem{Hyart2008}
T. Hyart, K. N. Alekseev, and E. V. Thuneberg,
Bloch gain in dc-ac-driven semiconductor superlattices in the absence of electric domains, 
Phys. Rev. B {\bf 77}, 165330 (2008).
\bibitem{Hyart2009a} 
T. Hyart, N. V. Alexeeva, J. Mattas, and K. N. Alekseev,
Terahertz Bloch Oscillator with a Modulated Bias,
Phys. Rev. Lett. {\bf 102}, 140405 (2009).
\bibitem{Hyart2009b} 
T. Hyart, J. Mattas, and K. N. Alekseev, 
Model of the Influence of an External Magnetic Field on the Gain of Terahertz Radiation from Semiconductor Superlattices, 
Phys. Rev. Lett. {\bf 103}, 117401 (2009).

\bibitem{Dahan1996}
M. Ben Dahan, E. Peik, J. Reichel, Y. Castin, and C. Salomon, 
Bloch Oscillations of Atoms in an Optical Potential, 
Phys. Rev. Lett. {\bf 76}, 4508 (1996).

\bibitem{Anderson1998}
B. P. Anderson, M. A. Kasevich, 
Macroscopic Quantum Interference from Atomic Tunnel Arrays, 
Science {\bf 282} (5394), 1686-1689 (1998).

\bibitem{Morsch2001}
O. Morsch, J. H. Muller, M. Cristiani, D. Ciampini, and E. Arimondo, 
Bloch Oscillations and Mean-Field Effects of Bose-Einstein Condensates in 1D Optical Lattices, Phys. Rev. Lett. {\bf 87}, 140402 (2001).

\bibitem{Cristiani2002}
M. Cristiani, O. Morsch, J. H. Muller, D. Ciampini, and E. Arimondo, 
Experimental properties of Bose-Einstein condensates in one-dimensional optical lattices: Bloch oscillations, Landau-Zener tunneling, and mean-field effects, 
Phys. Rev. A {\bf 65}, 063612 (2002).
\bibitem{Ott2004} H. Ott, E. de Mirandes, F. Ferlaino, G. Roati, G. Modugno, and M. Inguscio, 
Collisionally Induced Transport in Periodic Potentials, 
Phys. Rev. Lett. {\bf 92}, 160601 (2004).

%\bibitem{Gluck01}  
%M. Gluck, F. Keck, A. R. Kolovsky, and H. J. Korsch, Wannier-Stark States of a Quantum Particle in 2D Lattices, 
%Phys. Rev. Lett. {\bf 86}, 3116 (2001).
\bibitem{Gluck02}  
M. Gluck, F. Keck, A. R. Kolovsky, and H. J. Korsch, Wannier-Stark resonances in optical and semiconductor superlattices,
Phys. Reps. {\bf 366} (3), 103-182 (2002).
%States of a Quantum Particle in 2D Lattices, 
%Phys. Rev. Lett. 86, 3116 (2001).
\bibitem{Dmitriev01}  I. A. Dmitriev and R. A. Suris, Electron localization and bloch oscillations in quantum-dot superlattices under a constant electric field, Semiconductors {\bf 35}, 212 (2001).
\bibitem{Dmitriev02}  I. A. Dmitriev and R. A. Suris, Damping of Bloch oscillations in quantum dot superlattices: A general approach, Semiconductors {\bf 36}, 1364 (2002).
%\bibitem{Kolovsky2003}
%A. R. Kolovsky, H. J. Korsch, Phys Rev A 67, 063601 (2004)
\bibitem{Kolovsky2013} 
A. R. Kolovsky, E. N. Bulgakov,
Wannier-Stark states and Bloch oscillations in the honeycomb lattice,
Phys. Rev. A {\bf 87} (3), 033602 (2013).

\bibitem{Bistritzer2011}
R. Bistritzer and A. H. MacDonald, Moir\'e bands in twisted double-layer graphene. Proc. Nat. Acad. Sci. {\bf 108}, 12233-12237 (2011).
\bibitem{Cao2018a} Y. Cao, V. Fatemi,  A. Demir, S. Fang, S. L. Tomarken, J. Y. Luo, J. D. Sanchez-Yamagishi, K. Watanabe, T. Taniguchi, E. Kaxiras, R. C. Ashoori, and P. Jarillo-Herrero, Correlated insulator behaviour at half-filling in magic-angle graphene superlattices. Nature 556, 80-84 (2018).
\bibitem{Cao2018b} Y. Cao, V. Fatemi,  S. Fang, K. Watanabe, T. Taniguchi, E. Kaxiras, and P. Jarillo-Herrero, 
Unconventional superconductivity in magic-angle graphene superlattices. Nature 556, 43-50 (2018).
\bibitem{Cao2016} Y. Cao, J. Y. Luo, V. Fatemi,  S. Fang, J. D. Sanchez-Yamagishi, K. Watanabe, T. Taniguchi, E. Kaxiras, and P. Jarillo-Herrero, Superlattice-Induced Insulating States and Valley-Protected Orbits in Twisted Bilayer Graphene. 
Phys. Rev. Lett. {\bf 117}, 116804 (2016).
\bibitem{Kim2016} %DOI: 10.1021/acs.nanolett.6b01906
Y. Kim, P. Herlinger, P. Moon, M. Koshino, T. Taniguchi, K. Watanabe and J. H. Smet, 
Charge Inversion and Topological Phase Transition at a Twist Angle Induced van Hove Singularity of Bilayer Graphene, 
Nano Lett. {\bf 16}, 5053-5059 (2016). 
\bibitem{Berdyugin2020} A. I. Berdyugin, B. Tsim, P. Kumaravadivel, S. G. Xu, A. Ceferino, A. Knothe, R. Krishna Kumar, T. Taniguchi, K. Watanabe, A. K. Geim, I. V. Grigorieva, V. I. Fal’ko, 
Minibands in twisted bilayer graphene probed by magnetic focusing,
Sci. Adv. 6: eaay7838 (2020)
\bibitem{Bistritzer2009} %Bistritzer, R. and MacDonald, A. H., 
R. Bistritzer and  A. H. MacDonald, Electronic cooling in graphene, Phys. Rev. Lett., 102, 206410 (2009).
\bibitem{Tse2009} W. K. Tse, S. Das Sarma, Energy relaxation of hot Dirac
fermions in graphene. Phys. Rev. B. {\bf 79}, 235406 (2009).
\bibitem{Song2012} J. C. W. Song, M. Y. Reizer, L. S. Levitov, Disorder-assisted electron-phonon scattering and cooling pathways in graphene. Phys. Rev. Lett. {\bf 109}, 106602 (2012).
%\bibitem{zak} 
%J. Zak, Magnetic Bloch oscillations, Europhys. Lett. {\bf 105}, 67001 (2014)

\bibitem{Ju2011} L. Ju, et al. Graphene plasmonics for tunable terahertz metamaterials. Nat. Nanotechnol. 6, 630–643 (2011).

\bibitem{Yan2012a}  H. Yan, et al. Tunable infrared plasmonic devices using graphene/insulator stacks. Nat. Nanotechnol. {\bf 7}, 330-334 (2012).

\bibitem{Yan2012b}  H. Yan, et al. Infrared spectroscopy of tunable dirac terahertz magneto-plasmons in graphene. Nano. Lett. {\bf 12}, 3766-3771 (2012).

\bibitem{Tu2020} N. H. Tu, K. Yoshioka, S. Sasaki, M. Takamura, K. Muraki and N. Kumada, Active spatial control of terahertz plasmons in graphene, Communications Materials {\bf 1}:7 (2020). %, Article number: 7 (2020)

\bibitem{Ateshian2020} 
L. Ateshian, H. Choi, M. Heuck, and D. Englund,
Terahertz Light Sources by Electronic-Oscillator-Driven Second Harmonic Generation in Extreme-Confinement Cavities, \href{https://arxiv.org/pdf/2009.13029.pdf}{arXiv:2009.13029}
%LL \bibitem{kogan_shulman} Sh. M. Kogan and A. Ya. Shulman, Extraneous random forces and equations for correlation functions in the theory of nonequilibrium fluctuations, Fiz. Tverd. Tela (Leningrad) {\bf 12}(4), 1119-1123 (1970) [Sov. Phys. -- Solid State {\bf 12}(4), 874 (1970)].
%Sov. Phys. JETP 56, 862 (1969); Sh. M. Kogan and A. Ya. Shulman, Sov. Phys. JETP 57, 2112 (1969);
%LL \bibitem{kogan_book} Sh. Kogan, Electronic Noise and Fluctuations in Solids (Cambridge University Press, 2008)
%\bibitem{Stephen2019}
%\addLL{Stephen Carr, Shiang Fang, Ziyan Zhu, and Efthimios Kaxiras, arXiv:1901.03420}

%\bibitem{Denschlag}
%J Hecker Denschlag, J E Simsarian‡, H H¨affner†,
%C McKenzie, A Browaeys, D Cho§, K Helmerson,
%S L Rolston, and W D Phillips

\bibitem{Supplementary Information} %Online Supplement
See Supplementary Material for detailed estimates of phonon emission and carrier dephasing rates, and a step-by-step derivation of the backaction on the oscillator due to Bloch-oscillating carriers.

\end{thebibliography}
\end{document}